\ifdefined\ONECOLUMN
\documentclass[journal, 12pt, onecolumn, draftclsnofoot, letterpaper]{IEEEtran}
\else
\documentclass[journal, a4paper, twoside]{IEEEtran}
\fi

\IEEEoverridecommandlockouts
\usepackage{algorithm,algorithmic}
\usepackage{amsmath,amssymb,mathrsfs}
\usepackage{cite}
\usepackage{subfigure}
\usepackage{graphicx,color}
\usepackage[nomain,acronym,toc]{glossaries}
\usepackage{hyperref}
\usepackage{subfigure}
\usepackage{balance}
\usepackage{tikz,siunitx}
\usepackage{url}
\usepackage{graphicx} 
\usepackage{epstopdf}
\usepackage{multirow}
\usepackage{array}
\usepackage{amsmath,amssymb}
\usepackage{makecell}
\usepackage{setspace}
\usepackage{multirow}
\usepackage{booktabs}
\usepackage{tabularx}
\definecolor{sblue}{RGB}{0,51,200}
\definecolor{sred}{RGB}{200,51,130}

\renewcommand{\eqref}[1]{(\ref{#1})}

\ifCLASSINFOpdf
\else
\fi

\begin{document}
	\title{A Survey of Computation Offloading \\with Task Types}
	\author{Siqi Zhang, Na Yi and Yi Ma
		\thanks{The authors are with Institute for Communication Systems, University of Surrey, UK, GU2 7XH. e-mails:
			(s.zhang, n.yi, y.ma)@surrey.ac.uk. ({\it Corresponding author: Na Yi})
			
					 This work is supported by the UK Department for Science, Innovation and Technology under the Future Open Networks Research Challenge project TUDOR (Towards Ubiquitous 3D Open Resilient Network).
			}
		
		}

\markboth{IEEE Transactions on Intelligent Transportation Systems}
{Transactions on Intelligent Transportation Systems}

\IEEEaftertitletext{\vspace{-2.6\baselineskip}}
\maketitle	

\begin{abstract}\label{abs}
Computation task offloading plays a crucial role in facilitating computation-intensive applications and edge intelligence, particularly in response to the explosive growth of massive data generation.
Various enabling techniques, wireless technologies and mechanisms have already been proposed for task offloading, primarily aimed at improving the quality of services (QoS) for users. 
While there exists an extensive body of literature on this topic, exploring computation offloading from the standpoint of task types has been relatively underrepresented.
This motivates our survey, which seeks to classify the state-of-the-art (SoTA) from the task type point-of-view. 
To achieve this, a thorough literature review is conducted to reveal the SoTA from various aspects, including architecture, objective, offloading strategy, and task types, with the consideration of task generation. It has been observed that task types are associated with data and have an impact on the offloading process, including elements like resource allocation and task assignment.
Building upon this insight, computation offloading is categorized into two groups based on task types: static task-based offloading and dynamic task-based offloading.
Finally, a prospective view of the challenges and opportunities in the field of future computation offloading is presented. 
\end{abstract}


\section{Introduction}\label{Sec01}
Nowadays, with the rapid growth of computation-intensive and latency-sensitive mobile applications has significantly increased the demands on user equipment (UE) in terms of computation capacity and battery life (see \cite{6978603,7879258,7883826}). 
Despite considerable enhancements for UEs, they still fall short of meeting these demands. 
Local execution of these applications on UEs could result in a range of issues, including processing errors for computational tasks and task timeouts \cite{7488250,8985893,Jiang2019,9139976}. 

\begin{figure}[t]
\includegraphics[width=.5\textwidth]{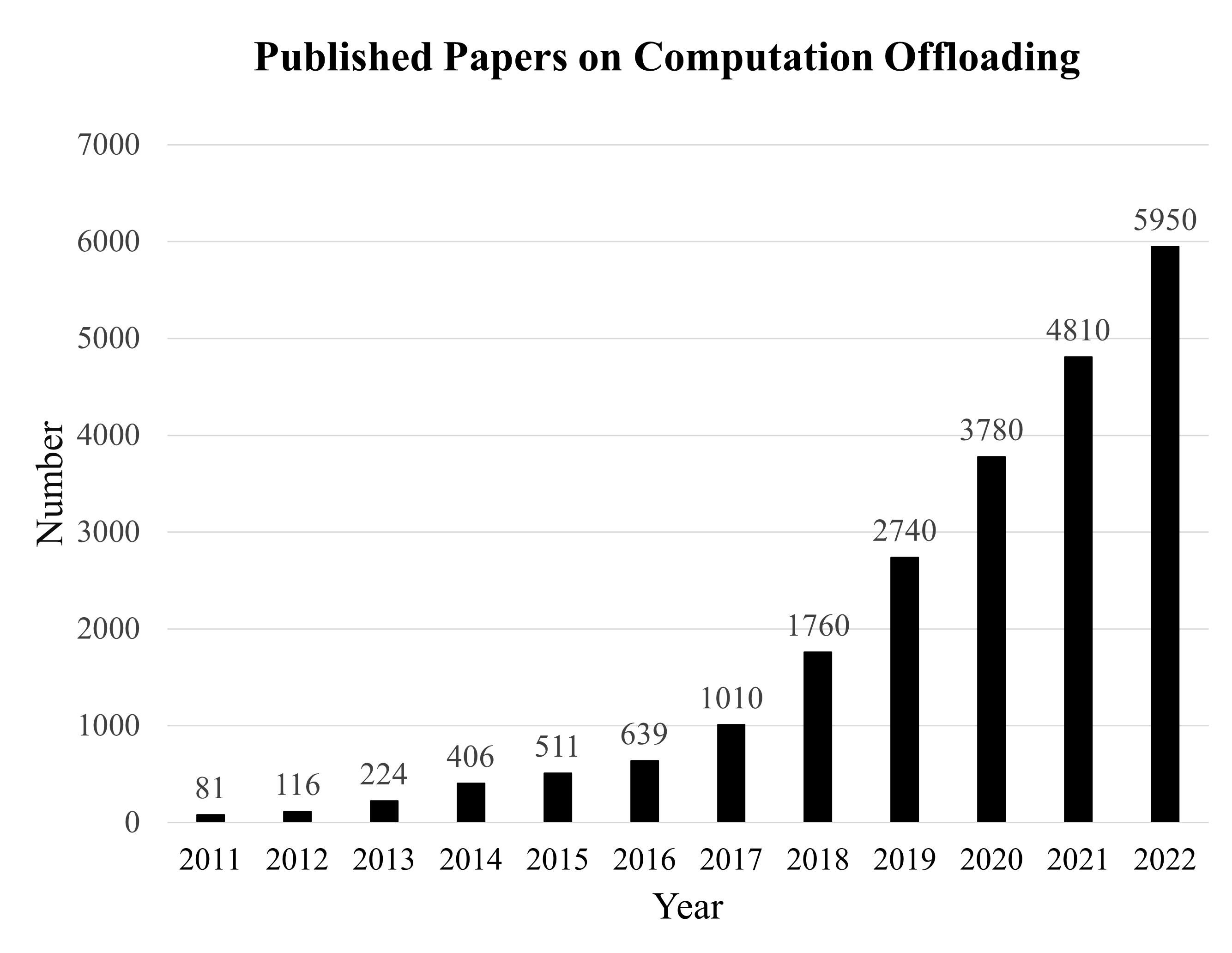}
\caption{ The number of computation offloading related publications per year from 2011 to 2022 (data collected from Google Scholar with key words ``computation offloading'').}
\label{Fig01}
\end{figure}

Computation offloading refers to the practice of relocating computational tasks from a device with limited resources to a more capable remote processing node (RPN), like a cluster, grid, cloud, or edge server \cite{7879258}.
Offloading the computation-intensive and latency-sensitive tasks from UE to RPNs with extra computation resource has been recognized as one of the feasible solutions to handle such issues mentioned above. 
The powerful computation resource of RPN can largely compensate the insufficient computation capability of the UE.
Therefore, more researchers are focusing on computation offloading. 
As depicted in Fig.~\ref{Fig01}, the recent years have witnessed a dramatic increase in the number of publications on computation offloading. These publications cover a broad spectrum of scenarios, ranging from general systems to specialized areas such as smart transportation systems, robotics swarms, smart home systems \cite{7879258}. This trend demonstrates the versatile application and growing interest in computation offloading across various fields.

\begin{figure*}[t]\centering
\begin{minipage}{0.49\linewidth}
\vspace{3pt}
\centerline{\includegraphics[scale=0.32]{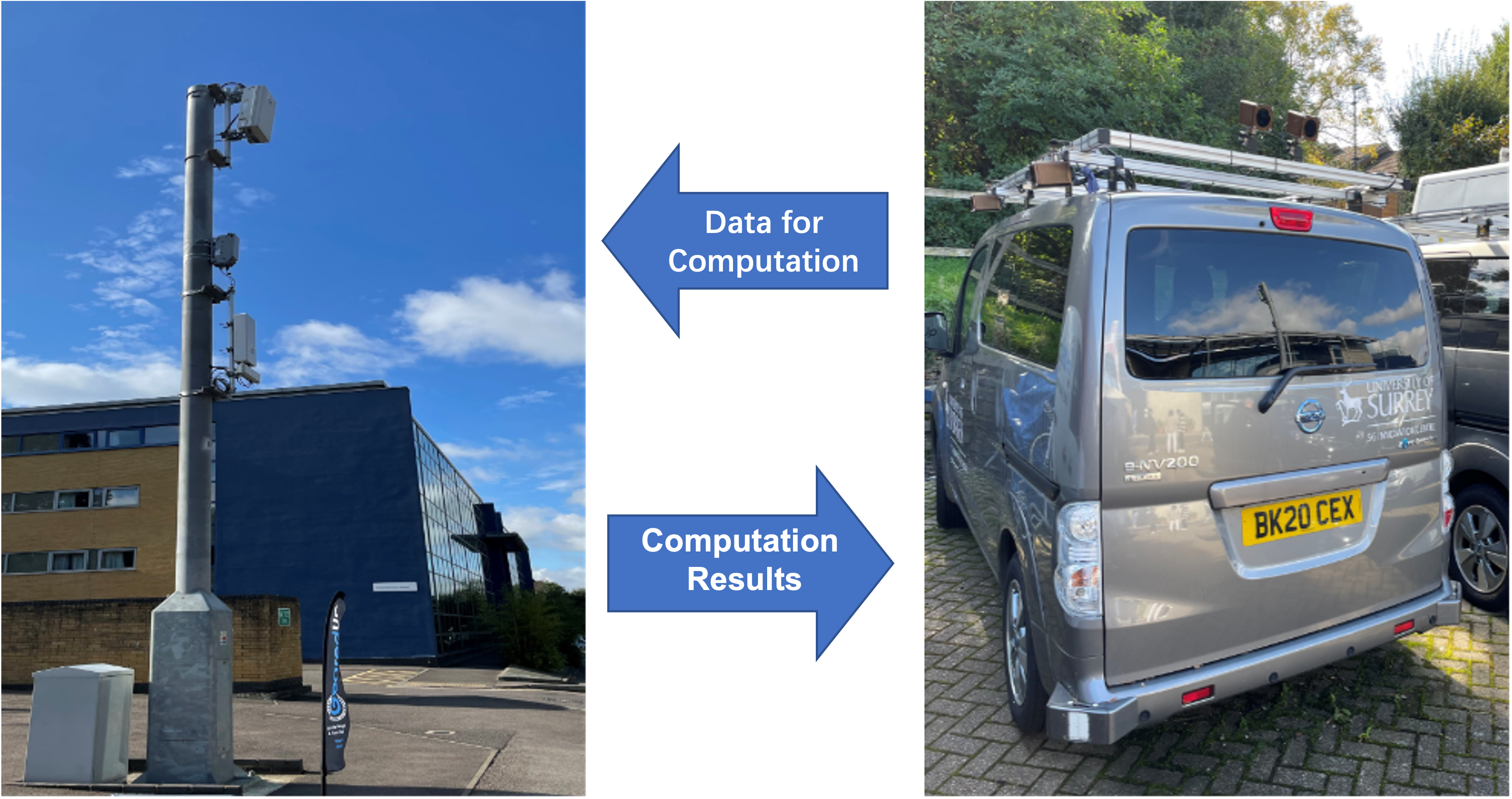}}
\centerline{Autonomous vehicle}
\end{minipage}
\quad
\begin{minipage}{0.48\linewidth}
\vspace{3pt}
\centerline{\includegraphics[scale=0.32]{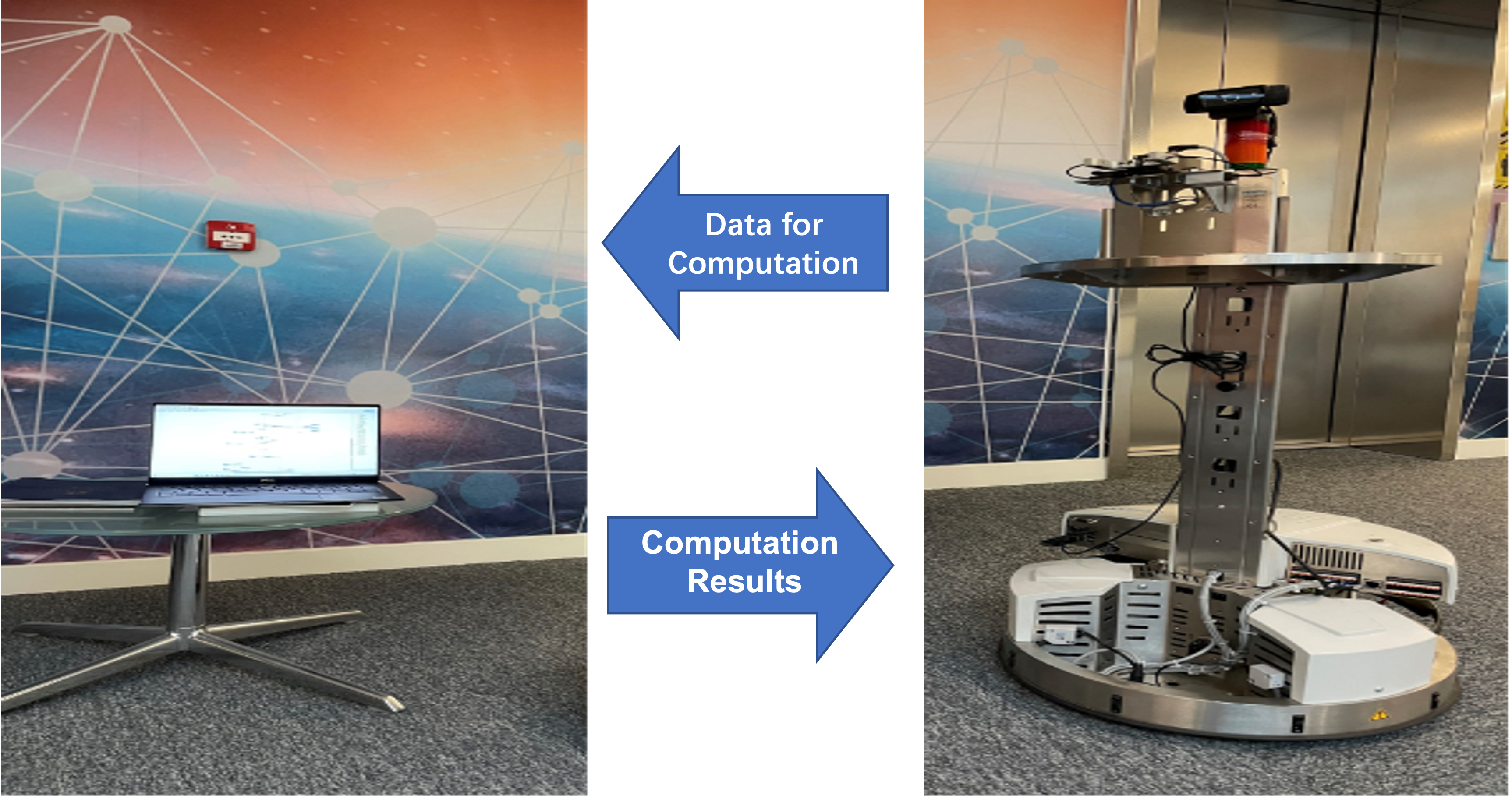}}
\centerline{Robot platform}
\end{minipage}
\caption{Examples of computation offloading applications (Provided by 5GIC \& 6GIC, University of Surrey).}
\label{Fig02}
\end{figure*}

Many research efforts have been paid to investigate computation offloading in various applications, i.e., autonomous vehicles \cite{9440722,9181432,9047920,9109630,9334447,9403985,9531562}, unmanned aerial vehicle (UAV) \cite{8654698,8445936,yu2020joint,8956055,9364753}, cloud gaming \cite{8648419,8761494,9120235}, and robot swarm \cite{8653406,9183527,9369456,8955967,7381995,8363064,zhang2023robot}. 
Examples of computation offloading applications are also shown in Fig.~\ref{Fig02}.	
The left panel of Fig~\ref{Fig02} illustrates an example of an autonomous vehicle. In this scenario,  the vehicle transmits sensor data to a nearby base station, which is linked to an edge server. This server analyzes the data and sends back the processed results to the vehicle through the base station, facilitating autonomous driving.
The right panel of Fig~\ref{Fig02} depicts an example of robot based task offloading. Here, the robot is responsible for sending sensor data to the computer. 
After processing this data, the computer generates control signals and sends them back to the robot.

Various challenges exist in the computation offloading across different scenarios.
One such challenge arises from the dynamic task arrival rate, often experienced in the autonomous vehicle scenarios.
This issue can hinder the efficient and effective allocation of resources during computation offloading. If not managed properly, it can lead to resource underutilization or overloading, which in turn can result in increased costs and potential system failures \cite{9440722}.
Another significant challenge in computation offloading research is the network congestion and considerable resource waste caused by transmitting large volumes of redundant data. 
This issue is particularly prominent in scenarios such as cloud gaming, robot swarms, and collaborative UAV applications \cite{2104.03818, zhang2023robot, 8653406}.

The architecture of computation offloading also received much attentions. 
Mobile cloud computing (MCC) was proposed to be used to help the computing task offloading in \cite{dinh2013survey}. 
However, cloud servers are generally far away from the UE. When UE needs to transmit the data to the cloud for processing the task, although the computation resource at the cloud is sufficient, the latency caused by transmission is considerable. 
Moreover, due to the characteristic of centralized processing of MCC, a large amount of data is offloaded from the UE to the cloud. 
This will cause excessive network overload. To solve this problem, another solution was proposed to move the RPN to the edge of the network.

One of the early schemes by bringing RPN closer to the UE was to use mobile edge computing/ multi-access edge computing (MEC)\cite{nguyen2020smartphone}. European Telecommunications Standards Institute (ETSI) released the MEC standardization in 2014 \cite{hu2015mobile}. 
The purpose of MEC is to move the computation to the edge of network and bring the RPN closer to the UE. 
It can achieve lower communication latency than MCC. Another scheme, fog computing (FC), was introduced in 2012 by Cisco to enable the processing of the applications on billions of connected devices at the edge of the network \cite{bonomi2012fog}. 
It uses edge devices which are closer to UE to process a large number of tasks, so that it is more suitable for the context of Internet of Things (IoT) by comparing with MEC. 
Moreover, such as cloud-fog, cloud-MEC, these multi-level computation offloading architectures were also often considered \cite{9217600}. 
These architectures can maximize the use of the advantages of different computing methods. 
Multi-level computation offloading was used when the number of tasks is large and the requirements (latency, computation capacity) of the tasks vary greatly \cite{9760072}.

Another focus of computation offloading research is the optimization, which mainly focused on task allocation and resource allocation~\cite{9403939}. 
To explore the diversity brought by fading channels \cite{720551,wang2020task,9459529} and multiple RPNs, two folds of work has been proposed to investigate task allocation and resource allocation correspondingly in \cite{9316648,9524604}. 
In multi-MEC scenario,  authors considered allocating different tasks to different MECs to achieve the best energy efficiency \cite{8895891,9316648}. 
In \cite{9165797,8279411}, authors considered splitting the task into subtasks and then allocating the subtask to different RPNs to gain energy efficiency.

For resource allocation, resources can be divided into two main categories: communication resource and computation resource \cite{9454395,9435770,9392259}. 
Bandwidth, communication time, signal frequency, relay nodes, and transmission power are considered as the main communication resource which need to be optimized in computation offloading to improve the performance \cite{huang2019deep,li2019offloading,9686591,9580302,9129054}. 
For computation resource, it includes the computation resources of UE and the computing resources of RPN. 
Computation resource allocation is common in multi-RPN scenarios and in partial offloading scenarios \cite{9500558,8552318}. 
In addition to these two major aspects, there are other aspects for various optimization methods and optimization objectives. 
For example, the splitting ratio needs to be taken into account when task splitting is used \cite{8279411}. 
Time allocation for energy-harvesting process needs to be studied while energy-harvesting is considered \cite{7442079}. 
Moreover, data correlation needs to be considered while optimizing the transmitted data \cite{9448864}.

It is also worth pointing out that the data generation of tasks has a significant impact on task offloading \cite{6963473}. 
According to the features of data generation, the tasks can be divided into two types in our viewpoint, i.e., static task and dynamic task (for detailed definition, please see Section \ref{Sec02A-1}). 
Dynamic task offloading often suffer higher complexity in optimization than static task offloading (e.g., dynamic changes in data size \cite{9448864}, dynamic arrival of tasks \cite{8269175}, dynamic computation density \cite{6574874}). 
Moreover, in some cases, static tasks can be converted into dynamic tasks through optimization. 
For example, in \cite{8279411}, authors converted the task from static task into dynamic task by considering the task splitting based on the channel state information (CSI) and UE computation capacity. 
In \cite{9448864}, authors considered the redundancy removal to dynamically remove part of the data to change the task from static task into dynamic task. Although the task type changing will increase the complexity of the computation offloading algorithm, it can improve the resource efficiency, i.e. energy efficiency.

To the best of our knowledge, there is no comprehensive survey has been provided according to the different types of task for computation offloading. 
Therefore, in this paper, we propose a novel classification method, and employ the historical method to survey the corresponding work over these two categories.
\begin{figure*}[t]
	\centering
	\includegraphics[scale=0.36]{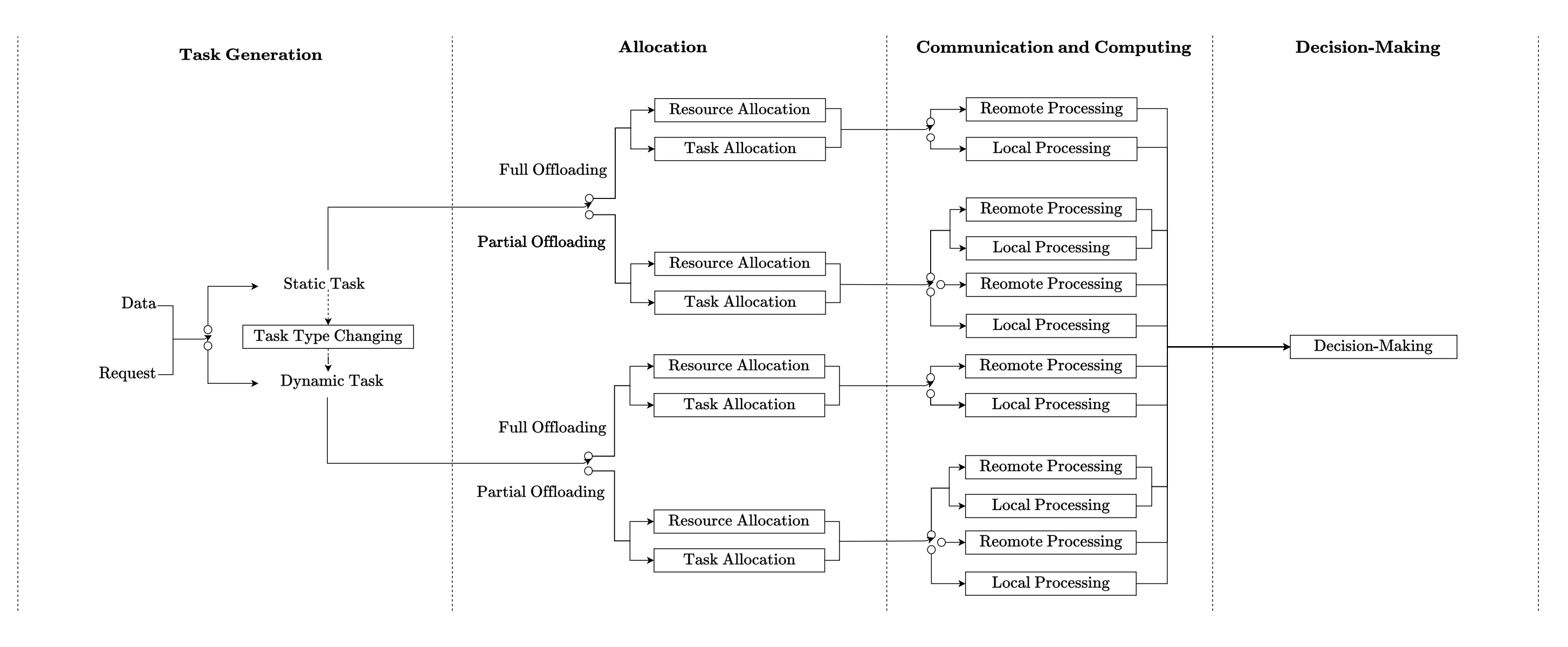}
	\caption{Computation offloading process.}
	\label{Fig03}
	\vspace{-1em}
\end{figure*}
The rest of this paper is organized as following: an overview of computation offloading is given in Section \ref{Sec02}, which includes the analysis of the computation offloading process. 
Next, the related survey papers about computation offloading are introduced in Section \ref{Sec031}.
Then, a detailed survey is conducted for computation offloading based on static task in Section \ref{Sec03} and dynamic task from energy perspective in Section \ref{Sec04} respectively. 
Finally, in Section \ref{Sec05}, we present a discussion, forecasting the possible evolution and the corresponding challenges that computation offloading will have in the years to come, and the conclusion.

\section{Overview of Computation Offloading}\label{Sec02}
The overview of computation offloading is given in this section. 
We review the computation offloading process including the current state-of-the-art (SoTA) and challenges in the different steps.
Then, we summarize the current hot directions in computation offloading related research.

\begin{table*}
	\centering
	\caption{Papers corresponding to each category}\label{tab01}
	\begin{tabularx}{\textwidth}{|cX|l|X|} 
		\hline
		\noalign{\hrule height 0.3pt}
		& & \thead{\textbf{\normalsize Static Task Generation}} & \thead{\textbf {\normalsize Dynamic Task Generation}}  \\ 
		\hline
		\multirow{7}{*}{\thead{\textbf{\normalsize Optimization}\\ \textbf{\normalsize Objective}}} 
		& Energy Saving Maximization & & \\ 
		\cline{2-4}
		& \quad 1) \quad Energy consumption of UE only& [61]-[66] & [29], [56], [57], [59], [60], [67]-[74] \\  
		\cline{2-4}
		& \quad 2) \quad Energy consumption of multiple entities (including various combinations of UE, relay node, edge or cloud server) &  [8], [19], [75]–[78] &  [79], [80] \\
		\cline{2-4}
		& Latency Minimization & \parbox{5cm}{[10], [22], [81]-[85]} &  [86]–[89] \\ 
		\cline{2-4}
		& Multi-objective (energy, latency, and reliability) &  [15], [17], [23], [44], [90]–[101] & [25], [45] \\ 
		
		\hline
		\multirow{2}{*}{\thead{\textbf{\normalsize Collaboration}\\ \textbf{\normalsize Strategy}}} 
		& Non-Collaboration & \parbox{5cm}{ [22], [61]–[65], [75], [76], [81],
			[83]–[85], [90]–[94], [96]–[103]} & \parbox{4cm}{ \vspace{0.3em}[16], [30], [56], [57], [59], [60], [67]-[74], [86]-[89], [104]–[107]} \\  
		\cline{2-4}
		
		& Collaboration & \parbox{5cm}{ \vspace{0.3em}[8], [15], [17], [19], [23], [43],
			[66], [77], [82], [95], [108]–[110]} &  [111]–[114] \\
		\hline
		\multirow{2}{*}{\thead{\textbf{\normalsize Offloading}\\ \textbf{\normalsize Strategy}}} 
		& Full Offloading & \parbox{5cm}{\vspace{0.3em}[8], [10], [15], [43], [63], [65],
			[75], [81], [82], [84], [91], [93],
			[96]–[102], [109]} & \parbox{4cm}{ \vspace{0.3em}[11], [16], [45], [56], [59],
			[60], [68], [71], [86], [104],
			[105]} \\ 
		\cline{2-4}
		& Partial Offloading & \parbox{4.5cm}{\vspace{0.3em}[44], [61], [62], [76], [77], [83], [85], [90], [92], [94], [103]} & \parbox{4.5cm}{\vspace{0.3em}[25], [67], [70], [72]–[74], [87], [88], [106], [107], [111]} \\
		\hline
		\multirow{4}{*}{\thead{\textbf{\normalsize Application}\\ \textbf{\normalsize Domain}}} 
		& IoT &  [62], [63], [88], [95], [96] &  [73] \\ 
		\cline{2-4}
		& Autonomous vehicle & [8], [19], [63], [66] &  [11], [69] \\
		\cline{2-4}
		& Robotics &  [23] &  [25], [29] \\
		\cline{2-4}
		& AR and VR &  [20], [22] & [21], [106] \\
		\hline
		\noalign{\hrule height 0.3pt}
	\end{tabularx}
\end{table*}

\subsection{Computation Offloading Process}\label{Sec02A}
In our work, the computation offloading process is divided into four steps: task generation, allocation, communication and computing, and decision making as shown in Fig. \ref{Fig03}.

\subsubsection{Task generation}\label{Sec02A-1}
The step one of computation offloading is task generation. 
In the computer science domain, a task is originally defined as the basic programming unit controlled by the operating system \cite{sheldon2023task}.
This definition is not suitable for the computation offloading system, since the relationship between the various factors formed the task is not described. 
Thus, based on \cite{sheldon2023task}, the definition of task is redefined as: a task is the basic programming unit controlled by the operating system, and it is formed by the request, data. 
Once a UE receives the request, it has to start using the corresponding functions\footnote{When function does not appear with objective (objective function), it refers to the function used to process data} to process the corresponding data (or offload the corresponding data). 
The generation of data can be divided into periodic and non-periodic generation, and generation of request can also be classified in this way. 
Thus, the task is only generated periodically if both data and request are generated periodically.

According to the relationship among tasks, data, and requests, tasks can be categorized as either dynamic or static. 
If the task of UE is static task, the following two conditions need to be met at the same time: tasks are generated periodically; the data size and CPU cycles requirement of the corresponding task in different time will not change. 
If the two conditions of the static task are not met simultaneously, the task of UE is defined as dynamic task.
Furthermore, based on the proposed concept of task type, dynamic tasks typically arise in two manners. 
The first is when the tasks are directly generated during the task generation process (referred to as the first type of dynamic task). 
The second is when dynamic tasks are derived by intervening the generated static tasks (referred to as the second type of dynamic task).
Common intervening methods include task splitting \cite{8638800, 8279411} and redundancy removing \cite{9448864}.
The task obtained by splitting a static task can be seen as a dynamic task, because the ratio of the splitting in different time slots may be different, dynamically changing the data size of each subtask.
Redundancy removing can also be assume as an effective intervention of changing static tasks into dynamic ones, this is because the data size for each task will vary according to the correlation degree.
Task generation plays a pivotal role within transportation systems \cite{9181432,	ismail2022qos,materwala2022energy,ismail2021escove,ismail2021iot}. 
In such systems, tasks might involve route planning for vehicles, traffic flow prediction, or accident detection. 
For instance, when an autonomous vehicle needs to navigate from one point to another \cite{9495832}, it generates a task to plan the optimal route. 
This task can take on a static form, with a predetermined route set before departure, or it can be dynamic, adapting based on real-time traffic updates and changes in the environment.
Furthermore, consider smart traffic signal systems in urban settings. 
These systems may generate tasks aimed at forecasting future traffic flow, allowing for more efficient adjustments in the timing of traffic lights.

In SoTA studies exploring computation offloading with static tasks \cite{9440722, 9047920, 10091508, 10163408} (for further references, see Table \ref{tab01}), certain parameters such as task generation frequency, data size, and their associated CPU cycles remain constant. 
This fixed state simplifies the optimization process by reducing its complexity.
Conversely, in SoTA studies on computation offloading with dynamic tasks \cite{9448864, 10092918, 10247107, 10246262} (for further references, see Table \ref{tab01}), due to the dynamic nature of these tasks, further optimizations are required to design the optimization function (e.g., optimization in queuing problems \cite{10092918}, task splitting ratios \cite{7348012}, and data removal ratios \cite{9448864}). 
Optimizing computation offloading with dynamic tasks offers greater flexibility but also introduces increased complexity.
Furthermore, converting a static task into a dynamic one through intervention significantly enhances its flexibility in resource and task allocation. However, this transformation also substantially amplifies its complexity.

Most of the computation offloading related research is based on the properties of the generated task (i.e., static task or  dynamic task) to directly perform resource allocation, task allocation, and other optimizations \cite{7879258}.  
Therefore, most of their optimization starts from the step two of computation offloading.
When changing the task type is taken into account, it can be assumed that the corresponding optimization starts from the first step of computation offloading.

\subsubsection{Allocation}\label{Sec02A-2}
The second step is allocation, and it includes the resource allocation and task allocation. 
This step is the focus of current computation offloading related research \cite{7879258}. 
We list some key points in Table \ref{tab01}, and introduce them in detailed in the following part.

Optimization objective is one of the important key points needed to carry out the allocation.
Different scenarios and applications have various restrictions and requirements for tasks, so their corresponding optimization objectives will not be the same. 
For example,  ultra-reliable low latency edge computation need to consider the reliability; energy consumption of cloud server in MCC will not be considered usually. 
Based on \cite{9217600, 6978603, 7879258}, the main optimization objectives are  summarized as follows.

The first optimization objective is energy saving maximization.  
The energy consumption is the focus of current research on computation offloading, and it includes energy consumption of the computing and communication \cite{7879258}. 
The energy consumption caused by computing mainly includes energy consumption of local computing and RPN computing. 
The energy consumption caused by communication is primarily composed of communication (downlink and uplink) between UE and relaying node \cite{8756083}, different UEs \cite{7907225}, UE and RPN \cite{FENG2022103366}. 
Moreover, when a UE is on standby or a task is in the queue, the UE also need to consume energy. 
It is called the energy consumption caused by waiting, and because this part of the energy consumption is tiny (comparing with energy consumption of  communication and computing), most researchers directly ignore it \cite{9217600}.

The optimization objects will be different for different scenarios and applications.
Typically, in scenarios where tasks are offloaded to MCC or MEC servers with a continuous energy supply, researchers do not consider the energy consumption of these servers, as noted in several studies \cite{6574874, 7225153, 7249412, 7524497, 7553459, 7762913, 7914660, 7938331, 8249536, 8269175, 8305608, 8325478, 8341501, 8438562, 8468240, 8588982, 8638800, 8756083, 8854118, 8859632, 8956055, 9032141}.
However, there are many research works that consider the energy of edge and cloud servers in the optimization~\cite{8795328,8865058,9197634,8367359}, this is because cloud service providers and data centers need to optimize their energy usage to reduce their environmental impact and lower operating costs.
Moreover, in some cases where fog computing is considered based on IoT network, because some computing nodes themselves do not have continuous energy supply, it is necessary to consider the energy consumption of fog nodes  \cite{8745530, 9399667, 10.1145/2333660.2333724}. 
In some scenarios where relaying node is used to help UE and RPN to communicate, the energy consumption of relaying node will also be taken into account, such as using the UAV worked as the relaying node \cite{8956055, chen2020multilevel}. 

In some scenarios, the data transmitted by the UE to the RPN is much larger than the result of the task received by the UE from the RPN \cite{7442079, 9295872, 7553459, 8058414}. 
Therefore, the energy consumption of downlink transmission is often disregarded in optimization processes \cite{7442079, 9295872, 7553459, 8058414, zhang2023robot,9217600,cong2020survey}.
However, there are some applications, such as AR, because UEs need to download a large amount of data from the RPN (RPN with content provider), the energy consumption of downlink is the main optimization object, so the corresponding energy consumption for such applications cannot be ignored \cite{9120235, 6697941}.

The second optimization objective is latency minimization \cite{6849257, 7541539, 7874147, 8451923, 8555711, 8772198, 8956076, 9194161, 9201170, 8487354, 8246720, 8664595, 8648419, 8761494, 9120235, 8956076, 8350444}. 
Although the viewpoint of our survey is energy consumption, the studies with optimization objective of latency minimization would also use energy consumption as a constraint for optimization, so it is also necessary to investigate it.
The latency mainly includes computation latency and communication latency \cite{8387798,sadatdiynov2022review}. 
The computation latency also includes the latency of local (UE) computing and remote (RPN) computing. 
Communication latency includes the latency of communication (downlink and uplink) between different UEs,  UE and relaying node, UE and RPN, RPN and RPN. 
When the data size of the feedback signal from RPN to UE is small, most studies do not take into account the time cost of downlink \cite{8314696, 8956076, 8350444,9500558,9878080,9531966,9685884}.

In addition to the two mainstream optimization objectives of energy saving maximization and latency minimization, there are many other optimization objectives that are being explored in computation offloading research. 
These optimization problems often involve multiple objectives and trade-offs between factors, with energy consumption being a critical factor that is often considered.
The most researched optimization problem for multiple objectives is the trade-off between latency and energy consumption \cite{7225153, 7348012, 7463066, 7524497, 7870694, 7914660, 7973020, 8234573, 8325478, 8438562, 8756083, 8859632, 8952620, 8956055, 9165797, 8533343, 8760529, 7510999, 7510999,9349766}. 
In their work, they weighted energy consumption and latency, and designed different algorithms to minimize the weighted sum.
As an extension to the issue of time and energy costs, other costs are considered, such as the cost of renting an RPN \cite{7996857,8654698}.

Furthermore, reliability is often an important factor to be considered in computation offloading. 
The considered reliability includes communication reliability and computing reliability \cite{8603743}. 
An unreliable computation offloading system is meaningless, and in some studies, researchers assume that the errors caused by communication and computing do not affect the results of the task \cite{9448864}, so they can ignore the issues caused by reliability.
However, for computation offloading that works in the ultra-reliable low latency scenario, the factor of reliability can not be ignored, so most research regards reliability as a constraint to limit optimization function in this scenario. 
Moreover, with increasing demands on reliability, more and more researchers are now considering the trade-off between reliability and energy consumption in computation offloading \cite{8279411,8603743}.

Collaboration strategy is another key point to carry out the allocation. 
Different collaboration strategies have different optimization ideas. 
There are two types of collaboration in computation offloading, they are communication collaboration (the collaboration for communication between different devices) \cite{9750981} and computing collaboration (the collaboration for computing between different devices) \cite{9217600, 8895891}. 
Thus, the collaboration strategy for computation offloading can be divided into three different categories, they are non-collaboration, computing collaboration and communication collaboration. 
Different collaboration strategies have their advantages and disadvantages based on their features.

The first collaboration strategy is non-collaboration. 
When there is no collaboration between different UEs and no collaboration between different PRNs, the corresponding collaboration strategy is called non-collaboration \cite{9120235, 7225153, 7249412, 7524497, 7938331,7914660, 7553459,8325478, 8438562, 8854118, 9032141,10.1145/2333660.2333724, 9295872,8350444}.

The second collaboration strategy is computing collaboration. Computing collaboration is divided into two categories, they are RPN computing collaboration \cite{8270634, 8756083, 8745530, 8895891} and UE computing collaboration \cite{8653406}.
RPN computing collaboration computation offloading refers to different RPNs collaboratively processing tasks offloaded by UE.	
When computation offloading works in a complex multi-user, multi-task system, it is likely that the task requirements will vary significantly. 
Some tasks may be highly computation-intensive and require significant computing resources, while others may have strict latency constraints and require real-time processing.
Tasks that are computationally intensive but do not demand on low latency are better suited to be offloaded to the cloud server for processing;
tasks that do not require much computation but require extremely low latency are better suited to be offloaded to MEC servers or fog nodes for processing;
tasks with high computation and low latency requirements are more suitable to be offloaded to MEC servers for processing. 
When the tasks with different requirements arrive at the same time, it is difficult to meet all requirements of different tasks if there is only one RPN.
Therefore, this motivate researchers to consider this computing collaboration computation offloading \cite{7879258}.
The most common architecture of computing collaboration computation offloading is MEC-cloud \cite{8270634, 8756083, 8745530, 8895891,9566768}. 
In \cite{8270634}, Guo {\em et al}. used MEC-cloud collaboration for IoT over the fiber wireless networks to achieve the goal of energy consumption minimization. 
In \cite{8745530}, Zhao {\em et al}. focused on the collaboration computation offloading in the vehicular network, which is based on cloud-assisted mobile edge computing. 
The authors maximize system utility in their study by considering the task and resource allocation problems under latency constraints. 
UE computing collaboration computation offloading refers to different UEs collaboratively process tasks such as \cite{8911219, 9277911, 9495832, 7415983}. 
This collaboration often occurs in vehicle scenario. 
Allocation of UE's idle computing resources is the focus of research \cite{qiu2002scheduling}

The third collaboration strategy is communication collaboration. Communication collaboration is also divided into two categories, they are communication collaboration between UEs \cite{8756083} and communication collaboration between RPNs \cite{7907225}. 
The communication collaboration between UEs  refers to the collaboration between different UEs to transmit the data collaboratively, and it also be called multi-hop collaboration \cite{9750981}.
Many studies on computation offloading assume that all UEs can connect directly to the RPN through the wireless network, which is referred to as single-hop computation offloading \cite{8756083, 8370750}.
However, in practice, UE may experience network connectivity issues, even out of the coverage of RPN, and that may not be able to directly connect to the RPN. Thus, in some cases,  some UEs also need to act as relay nodes to help other UEs to transmit their data to the RPN, and it is called communication collaboration. 
For example, when the deployment of UEs is very dispersed, some UEs will be very close to the PRN and some UEs will be very far away from the PRN (even beyond the communication range). 
In this case, if single-hop communication is used, there is a high probability that the computation offloading of those UEs that are very far from the RPN will fail, leading to the failure of the computation offloading system. 
Therefore, this motivate researchers to consider the communication collaboration between UEs, those UEs close to the PRN need to act as relay nodes to help those UEs far from the PRN to offload their tasks in addition to offloading their own tasks \cite{7907225, 7510809, 8606972, 8300317, 8653406, 8956055, 9279235, 8653406, 8939441}. These studies involve a lot of relay node (UE) selection problems.
Communication collaboration between between UEs is often used in the scenarios involving  UAVs, robots, vehicles \cite{9364753, 8653406, 9183527, 8654698, yu2020joint, chen2020multilevel, 7572068}. 
Funai {\em et al}. \cite{8606972} studied trade-off problem in terms of computation delay and network lifetime in a cooperative multi-hop ad hoc network. Müller {\em et al}. \cite{7343576} investigated the problem of minimizing energy consumption for computation offloading in multi-hop wireless network.
Hong {\em et al}. \cite{8653406} studied the communication collaboration between different robots in robot swarms.  
The communication collaboration between RPNs  refers to the collaboration between different PRNs to transmit the data collaboratively \cite{7907225}. 
In the context of communication collaboration between RPNs, researchers focus on the problem of RPN selection for data transmission, i.e., which RPNs are used for the communication (as relay node) so that the processed data can be received by the moving UE \cite{7907225}.

Offloading strategy is the third point to carry out the allocation. 
The offloading strategy can be divided into two types, full offloading and partial offloading.

In full offloading, there will be two situations. 
The first situation of full offloading is that all tasks will be offloaded to the RPN for processing.  
It means that all the tasks need to be offloaded whatever the environment (wireless channel conditions, computation resource availability) changes, such as \cite{8279411,8269175,8664595}. 
This kind of situation is often considered in the sensor network because of the limited computation capacity of the local device. 
The second situation of full offloading is that UE offloads all the tasks to the RPN, or process all the task locally \cite{7307234,7553459,7938331}.
It means that all the tasks (generated in the same time slot), which belong the same UE, will be processed in the same terminal together.
Furthermore, when the UE only has one task and does not consider task splitting \cite{7553459}, then there are only two possible outcomes: either offload the task or not offload it.
Thus,  this case is identified as second situation of full offloading based on the definition that mentioned earlier.

In partial offloading \cite{7225153, 7249412, 7914660, 8438562, 8854118, 10.1145/2333660.2333724,9369456, 7762913, 8468240, 8638800, 8555711, 8956076}, UE can dynamically determine which tasks (for multi-task cases) or which part of task (if task splitting is considered) are offloaded to the RPN based on the channel state information, computation capability of UE, the computation capability of the RPN.
Unlike the second case in full offload, tasks of the same UE in the same time slot can be processed separately in different places. 
Moreover, considering the partial offloading with task splitting from application model. The applications can be divided into two types. 
They are data partitioned oriented application and code partitioned oriented application \cite{7879258}. 
For data partitioned oriented application, the corresponding tasks can all be divided into two parts of any size \cite{7448613,6963473}. 
On contrast, for code partitioned oriented application, the corresponding task can not be split into any size \cite{8885778}, and it requires the selection of the code to be offloaded (the data for the same code can not be separated) \cite{7060486}. 
Thus, data partitioning oriented application is more flexible than code partitioning oriented application. 
Moreover, considering the partial offloading with knowledge on the amount of data to be processed. 
The applications can be divided into two types. 
They are data amount known based application and  continuous-execution application \cite{7879258}. 
For the data amount known based application (such as the studies in \cite{8885778}), the amount of data which is need by the application is known before offloading process.  
For the  continuous-execution application (such as the studies in \cite{6413270,7136330}), it represents the application that the amount of data which is need by the application is unknown before offloading and  it is hard to estimate the data amount requirement. 
Thus, it will largely increase the difficulty of partial offloading comparing with the data amount known based application.

With the same optimization objective and collaboration strategy.  
Full offloading has a lower complexity in step two of computation offloading process \cite{7879258}, since in the first situation of full offloading, it is only necessary to consider offload all the task to RPN; in the second situation of full offloading, it is only necessary to consider the binary offloading further  (i.e. to offload or not to offload). 
However, in partial offloading, researchers need to study which tasks of the UE need to be offloaded, which tasks need to be executed locally, whether tasks need to be offloaded to different RPNs. 
In general, the optimization algorithm for full offloading is simpler, but its performance is worse. 
The optimization algorithm for partial offloading is a bit more complex, but the corresponding performance is better \cite{7879258}.

\subsubsection{Communication and computing}\label{Sec02A-3}
The third step of computation offloading is communication and computing. 
The step three corresponds to the data transmission process and the data processing process. In some cases, there are other processes, such as  energy-harvesting (EH) process. This step is actually the implementation of the optimization designed in step two and it is  just listed here to ensure the integrity of the computation offloading process. Thus, in the following, the survey for this step will not be carried out.

\subsubsection{Decision making}	\label{Sec02A-4}
After computing, UE needs to collect the task output (collect the results from UE itself, or download the results, such as the control signal, from RPN) to make the corresponding execution decision. 
Thus, there comes the step four, decision making. 
This step may involve task fusion or data fusion (e.g., in some cases where task splitting is considered, further data fusion and task fusion are required to get the final complete results).
This step is generally not taken into account by researchers in the optimization process.
It is just listed this step here to ensure the integrity of the computation offloading process, and  the survey for this step will not be carried out in the following part.

From our survey results, most researches on computation offloading focus on the second step, i.e., allocation (resource allocation and task allocation). 
Recently, more and more studies focus on joint optimization of step one and step two (e.g., joint optimization of task splitting and resource allocation and task allocation,  joint optimization of redundancy removing and resource allocation and task allocation). 
Therefore, in the following part, the work of these two steps will be surveyed in detail.

\subsection{Application of Computation Offloading}
Computation offloading has emerged as a pivotal technique in the realm of modern computing, especially in scenarios where devices have limited computational resources or energy constraints. Essentially, it involves transferring specific computational tasks from resource-constrained devices, such as smartphones or IoT devices, to more robust servers or cloud platforms. The primary objective is to enhance performance, conserve energy, and extend the battery life of the initiating devices.
In the context of mobile cloud computing \cite{7524497}, smartphones and tablets, which often grapple with limited battery life and computational capabilities, can greatly benefit from offloading. Intensive tasks, such as video processing, augmented reality applications, and high-end gaming, can be seamlessly transferred to the cloud, ensuring smoother user experiences without draining the device's resources.

The IoT is another domain where computation offloading proves particularly beneficial \cite{ismail2021iot}. For instance, in smart homes, devices like thermostats and security cameras can offload their data processing tasks to the cloud. 
This enables more efficient collaboration among various devices.
Similarly, in the healthcare sector, wearable devices like heart rate monitors or glucose meters can offload their data for more sophisticated analysis in the cloud, enabling real-time feedback and potentially life-saving alerts.

Autonomous vehicles \cite{materwala2022energy}, equipped with a plethora of sensors, stand to gain significantly from offloading. 
By transferring data to edge servers or cloud platforms, these vehicles can benefit from real-time processing, which is crucial for safe navigation and decision-making on the roads.

In the field of robotics \cite{zhang2023robot}, robots operating in dynamic environments can offload complex computational tasks, such as image recognition or path planning, to more powerful servers. This offloading not only facilitates their adept response to environmental changes but also enhances coordination among different robots.

Augmented reality (AR) and virtual reality (VR) \cite{9120235}, which demand high computational power for real-time rendering, can offload these intensive rendering tasks to the cloud. This ensures that users enjoy a smooth and immersive experience without overburdening the device's native hardware.

\subsection{Several Research Directions Related to Computation Offloading } \label{Sec02B}
In this subsection, several research directions related to the step one and the step two of computation offloading are introduce respectively. 
The current research directions on computation offloading with their corresponding papers are listed in Table \ref{structure2}, and it is used to summarize the problems in different research directions.

\subsubsection{Task Splitting}\label{Sec02B-1}
Task splitting is an important way to change the task type in the first step of computation offloading, and it can be jointly optimized with allocation (step two) to improve the performance of computation offloading.

Task splitting allows a task to be divided into multiple subtasks that can be allocated to different computation resources for parallel processing, which can reduce the overall execution time and energy consumption. 
By jointly optimizing task splitting and allocation, the system can make use of both local and remote resources effectively to achieve better performance. For example, a computationally intensive task can be split into two subtasks, with one subtask executed locally on the UE and the other offloaded to an RPN, in order to balance the workload and reduce the overall latency and energy consumption.

Task splitting is a approach to improve the energy efficiency of computation offloading \cite{8279411, 9165797, 8956076, 8603743, 7463066, 7249412, 7503859, 7874147}. 
It can change static tasks into dynamic tasks through splitting, so as to carry out more flexible resource allocation and task allocation. 
By using task splitting, tasks can be split into multiple subtasks\footnote{In the whole paper, subtasks are split by tasks}, which can be allocated to multiple RPNs for processing. 
It can improve the utilization of communication resources and computing resources to achieve energy-saving or latency reduction. 
There are two existing task splitting methods for computation offloading systems.
The first method is to split the task at any ratio with the assumption that there is full granularity in data partition of the task.
The second split method is to split the task by using code/source partitioning. 
This means that task splitting is no longer arbitrary (not in any ratio anymore) and  it has a certain splitting ratio (based on the attribute of code/source). 
In \cite{8279411,9165797,8956076}, the authors gave the assumption that there is full granularity in data partition. 
Thus, the considered task could be partitioned into subtasks of any size. 
However, this way of splitting is idealistic, so that in \cite{8603743, 7463066, 7249412, 7503859, 7874147}, the authors split the task according to the topological structure of task, such as code, function.

Two different task splitting methods have been introduced. 
In addition, task splitting can also be classified according to the number of subtasks after splitting, and there are two different types, they are splitting one task into two subtasks and splitting one task into more than two subtasks.
When considering splitting a task into two subtasks \cite{8638800}, it is common to consider that one subtask will be processed on the UE  and one subtask will be processed on the RPN. 
In this consideration, a UE only need to offload one of the subtasks to RPN during an offloading process, making split-ratio optimization a critical aspect of the optimization problem.
If the task is split into multiple subtasks \cite{8279411}, there is a more complex problem of multi-RPN node selection involved, in addition to determining the split ratio. 
Although this is more complexity, it provides greater flexibility and can lead to better performance.

According to the relationship between subtasks, task splitting  can also be divided into two types, i.e., parallel splitting and serial splitting. 
The consideration in \cite{9448864} is parallel splitting, and the subtasks after splitting can be processed serially or in parallel. 
In addition, authors \cite{8279411} considered serial splitting, and subtasks must be processed in a fixed order according to the topology of the corresponding task. 
The consideration in \cite{8603743} is mixed splitting, which combines parallel and serial splitting.

Different tasks/subtasks may be independent of each other or related to each other, as well as the subtask. 
When they are completely independent of each other, they can be processed in parallel, and there is no need to consider the sequential problem. 
When they are related to each other, it will involve the problem that the optimal offloading path selection is restricted by the topological structure of the task \cite{7249412}. 
Such issues often appear in the allocation of subtasks.

\subsubsection{Redundancy Removing for Computation Offloading}\label{Sec02B-2}
Redundancy removing is another  way to change the task type in the first step of computation offloading.
The data considered in computation offloading can be divided into two categories: correlated data and independent data. 
When considering correlated data, it means that the loss of certain data does not affect the processing of the corresponding task \cite{2103.15924, zhang2023robot}. 
Therefore, removing this redundant data can help save energy or reduce latency, without affecting the correctness of decision making.

There are some studies about redundancy removing in computation offloading from the perspective of input data. 
In Section \ref{Sec02B-1}, it is mentioned that different tasks or subtasks may be related to each other, the input data of the same task in the time domain may have a high correlation. 
Processing a task is actually an observation of the environment corresponding to the task. 
Under normal circumstances, the frequency of this observation is much greater than the change of the environment, which means that most observations are meaningless. 
This implies that the corresponding processing of the task is meaningless. 
Moreover, a task may be composed of multiple subtasks, so it may involve multiple observations of an environment from different angles. 
This kind of multiple observations of the same environment is also a kind of redundancy that can be removed. 
Nour {\em et al}. \cite{2103.15924} designed an experiment for object detection service based on the standard image dataset, ImageNet \cite{russakovsky2015imagenet}. The experiment proves that this kind of redundancy exists in the real world, which provides a theoretical basis for the optimization method based on redundancy removing.  

How to remove  redundancy to reduce the repeated data transmission and  repeated computation (include RPN computation and local computation) processes has become a new research hotspot \cite{10.1145/2333660.2333724}. 
In \cite{9448864}, Zhang {\em et al}. suggested that in the time domain, the input data of the same task at different times is repetitive on a large scale. 
Moreover, \cite{10.1145/2811587.2811598} and \cite{2104.03818} conducted the similar work. 
The task splitting mentioned in Section \ref{Sec02B-1} is a good way to be further explored to search and remove the overlap between different tasks \cite{9448864}. 
The work for redundancy removing in computation offloading is still relatively limited, but the performance improvement from the simulation results of existing papers is quite significant. 
As the specific work based on \cite{9448864, 2103.15924}, Nour {\em et al}. \cite{9397772} proposed  an efficient computing reuse architecture for edge computing called CoxNet. 
It enables the edge server to reuse the previous results while scheduling dependent incoming computing. 
Through evaluation based on real data sets, CoxNet can reduce task execution time by up to 50\%. 
Furthermore, Nour {\em et al}. considered this reuse architecture for IoT application \cite{2104.03818}.

Task splitting and redundancy removing are two ways to optimize computation offloading at the first step. 
Task splitting is initially designed to provide greater flexibility in resource allocation and task allocation, thereby improving system performance. 
Data has a great impact \cite{9397772} on computation offloading, but task splitting does not make good use of it. 
Redundancy removing  is an approach to optimize tasks by exploiting data correlation in the task generation process. 
It uses the huge impact of data, but it still receives little attention by researchers. In the following parts of this subsection,  the optimizations based on the second step of computation offloading are investigated.

\subsubsection{Ultra-Reliable Low-Latency Computation Offloading}\label{Sec02B-3}
Ultra-reliable low-latency edge computing is the new service for applications that demand high reliability and low latency, such as automation vehicle, industrial automation, and remote surgery. 
The ultra reliable low latency computation offloading has also attracted widespread attention, as seen in \cite{9186655, 8742606, 8734753, 8863420, 8555711, 9146125, 8911219, 8279411}. 
The conventional computation offloading system is designed based on average-based indicators, which can not meet the requirements of ultra reliability and low latency. 
As a result, changing the conventional average-based design architecture and taking ultra-reliable and low-latency into account for optimization presents a major challenge for computation offloading.
To address this challenge, some researchers considered task-queuing problem with dynamic task (task arrival rate can be assumed as a random variable) \cite{9186655, 8742606,  8734753, 8863420, 8555711, 9146125, 8911219}.  
Liu {\em et al.} \cite{8638800} considered the queuing problem with the reliability constraints, and their optimization objective is to minimize the energy consumption.
They modelled the latency and reliability constraints by using task queue lengths based on the extreme value theory. 
The above work considered the queue problem, which always has a close relationship with the different arrival rates. 
This indicates that the research direction of task-queuing for ultra reliable low latency computation offloading focuses on dynamic tasks.

The above research primarily  focuses on ultra reliable and low latency for the computation process. 
There is also a kind of research direction focusing the communication process for ultra reliable low latency computation offloading, so there is no excessive requirement on the task type. 
For example, Liu and Zhang \cite{8279411} considered the block error rate as the communication error, and a threshold for communication error was set up to constrain the communication error of the computation offloading system. 
Several research directions mentioned above (Section \ref{Sec02B-1}, \ref{Sec02B-2} and \ref{Sec02B-3}) are more aimed at dynamic tasks or the changing from static task to dynamic task, and they are more targeted. 
There are also some research directions that do not emphasize the task type too much as follows.

\subsubsection{Artificial intelligence (AI) Used for Computation Offloading}\label{Sec02B-4}
When considering task allocation and resource allocation in computation offloading, a lot of the issues that need to be considered are NP-hard problems \cite{8884234,9079564}.
For conventional computation offloading, many researchers try to solve them using heuristics \cite{8279411}, game theory \cite{8249536}. However, these approaches are less flexible and rely on a specific environment for optimization.
As a result, when the environment of the computation offloading system changes, the approaches may not achieve the optimum performance.
Since AI methods can learn the near optimum response strategies for different situations from existing data, such methods can more effectively solve complex offloading decision-making and highly dynamic problems.

AI is used in computation offloading systems to determine resource and task allocation.
It can work with both dynamic and static tasks and is used in the second step of computation offloading, without changing the type of task.
The current AI technology has wide applicability, and the introduction of AI into computation offloading systems is considered a solution to the aforementioned problems \cite{8884234,8270639,9109630,ranadheera2017mobile, 8713801, 8843978}.

Reinforcement learning (RL) is commonly used in computation offloading.
RL is a method of learning in dynamic systems that adjusts decisions based on whether the decision is positive or negative in different situations.
It sets up a reward and punishment mechanism.
Through continuous testing, it rewards and punishes a series of actions and then modifies its strategy.
After such continuous adjustments, the UE can learn which actions should be chosen in order to achieve the best return in certain situations, as in \cite{ranadheera2017mobile, 8713801, 8843978}.
In addition, there are Q-learning \cite{9181432}, deep reinforcement learning (DRL) \cite{8771176, 10.1145/3317572, 9161406}, and other techniques.

Beyond traditional approaches and reinforcement learning, meta-heuristic AI algorithms present another promising direction in the field of computation offloading. Meta-heuristic methods \cite{mishra2021review}, inherently designed to find approximate solutions for complex optimization problems, align seamlessly with the challenges of computation offloading, especially when dealing with NP-hard problems. Infusing AI techniques into these algorithms enables more adaptability and efficiency.
For instance, AI-augmented genetic algorithms \cite{ismail2022qos} can dynamically adapt crossover and mutation rates based on the observed performance of offloading decisions. Similarly, AI-powered simulated annealing can use machine learning models to ensure a better convergence. 
One significant advantage of combining AI with meta-heuristics is the ability to generalize across various offloading scenarios. Unlike traditional methods that might be tailored to specific environments, AI-enhanced meta-heuristics learn from a wide range of data, enabling them to perform effectively even when system conditions change.
Furthermore, with the vast amounts of data generated in computation offloading scenarios, deep learning models can be trained to identify patterns and insights that might be non-intuitive or too complex for traditional algorithms. This facilitates more intelligent initialization of meta-heuristics or even hybrid approaches that blend the strengths of various algorithms, guided by the insights of deep learning.

\begin{table*}[t]
	\centering
	\newcommand{\tabincell}[2]{\begin{tabular}{@{}#1@{}}#2\end{tabular}}
	\caption{Research on Computation Offloading with Corresponding Papers}\label{structure2}
	\begin{tabular}{|l|l|}	
		\hline
		\noalign{\hrule height 0.4pt}	
		\multicolumn{1}{|c|}{\textbf{Research Directions on Computation Offloading}}&\multicolumn{1}{c|}{\textbf{ Corresponding Papers}}\\	
		\hline	
		Energy-Harvesting (EH) for Battery Lifetime  &  [56], [65], [104], [105], [184]	\\
		\hline	
		Task Splitting for Allocation Flexibility &  [16], [44], [45], [57], [67], [72], [76], [88], [106], [125]\\
		\hline
		Redundancy Removing for Data Efficiency &  [30], [57], [89], [168]\\
		\hline
		Caching for Workload Reduction&  [11], [68], [71], [101], [114], [185]\\
		\hline
		Ultra Reliable Low Latency Computation Offloading &  [45], [59], [69], [72], [73], [87], [106], [111], [173]\\
		\hline
		Mobility Problem for Vehicle &  [8], [25], [66], [137], [186]–[188]\\
		\hline
		AI for Offloading-Decision Making
		&  [9], [11], [25], [73], [96], [105], [130], [177], [178] [175]
		\\
		\hline
		Collaboration Problem for UAV and Robot Swarm
		&   [15], [17], [19], [23], [24], [29], [131], [160]
		\\
		\hline
		\noalign{\hrule height 0.4pt}
	\end{tabular}
\end{table*}

\subsubsection{EH Used in Computation Offloading}\label{Sec02B-5}
Although there are many computation offloading algorithms for improving energy efficiency, which has maximized energy efficiency to a large extent, the process of computation offloading still requires energy and the problem of energy supply remains unsolved. 
Computation offloading need to be stopped when the battery of UE is used up. 
This can be overcome by charging the battery or using a larger battery. 
However, using the larger battery on mobile devices means increasing hardware costs, which is undesirable. 
Moreover, it may not even be possible to charge the battery of the UE in some application scenarios. EH is a promising technology to solve these problems. 
It can capture environmentally recyclable energy \cite{5522465}, including solar energy, wind energy. 
EH, as a way of energy supply, can  be used for the both static task and dynamic task.
Its resource allocation process is a bit more complicated, because in most of consideration, communication and EH cannot be done at the same time, so time resources need to be allocated.
To use this EH technology, Mao {\em et al}. \cite{7572018} established a EH model for computation offloading. 
In their model, the EH process is modeled as a continuous energy packet arrives, and this is a basic way of EH.  

You {\em et al}. \cite{7442079} presented a different EH method, microwave power transfer (MPT), in computation offloading system. 
The base station (BS) transfers power wirelessly to UE, and UE’s power comes entirely from MPT. 
In their work, authors assumed that local computing and MPT can work at the same time. 
However, the data transmission process and MPT can not work at the same time (work in half-duplex transmission). 
Because the energy conversion efficiency of this MPT is not high, the energy obtained by the UE from the MPT is also very low. 
Therefore, the authors also mentioned in their paper that this EH method is more suitable for low-complexity devices such as wearable computing devices and sensors.

\subsubsection{Mobility Problem in Computation Offloading}\label{Sec02B-6}
Mobility is one of the important features that can not be ignored in computation offloading \cite{8485780}, especially in the computation offloading for vehicle \cite{8676263}. 
This is therefore a feature that needs  to be taken into account, whether the task is static or dynamic.
Because of the mobility of UE, intermittent connections between UE and BS have become an often occurred state which can cause computation offloading to fail.   
According to our survey results, the mobility research in the field of computation offloading mainly focuses on task migration, and it includes selecting the appropriate RPN for offloading, and selecting the appropriate path for task (input data/output data) migration \cite{7045578, 8485780, 8255040, 6877640, 8676263}.

The location of the UE usually determines which RPN the UE needs to offload its task to, because the task is usually offloaded to the nearest RPN to achieve low latency requirement.  
Due to the limited coverage of the BS, the following situation may arise.
A UE transmits data to an RPN, after the transmission, if the UE is not within the coverage of the BS corresponding to this RPN, the UE will not receive the results corresponding to the transmitted data.
This situation often occurs in the vehicle-to-everything (V2X) environment \cite{8676263}, because the coverage of road side unit and BS is not large, but the moving speed of the vehicle is very fast.
Zhang {\em et al}. \cite{7907225} proposed a predictive combination-mode relegation scheme for MEC computation offloading in a cloud-enabled vehicle network. 
In order to solve the problem of UE leaving the coverage area of the previous corresponding MEC due to mobility, the authors provided two approaches based on the movement prediction. 
The movement of UE is unpredictable and irregular in most complex environments. 
Due to the development of machine learning, it is now possible to predict a UE's estimated stay time in a given area and its movement habits.
This provides a foundation for the development of this kind of computation offloading based on movement prediction~\cite{8676263}.

\subsubsection{Caching Used in Computation Offloading}\label{Sec02B-7}
The increasing demand for massive multimedia services over the mobile cellular network makes significant challenges to network capacity and backhaul links. 
The emergence of mobile edge caching and delivery techniques are promising solutions to cope with those challenges  \cite{6736753}.

In conventional centralized mobile network architectures, it is typical for remote internet content providers to supply UE with necessary input data for tasks, a scenario frequently encountered in VR and AR \cite{8305608}
In scenarios where numerous UEs require identical input data or the same UE repeatedly requests the same data, the mobile network is burdened with transmitting redundant data. This repetition can lead to substantial network congestion and an inefficient use of network resources.
Caching popular content at the edge of the network (it is not the Internet content provider, and it is close to UE) can cancel the repeated transmissions of the same content, which will significantly reduce the workload of communication and reduce energy consumption and decrease latency. 
In this case, the input data used to form the task does not come from the UE but the remote content provider. 
Another situation is that the UE itself generates data and the corresponding task, when the result of  a task is cached in MEC server, the UE can  download the result directly from the edge server \cite{ 8588982, 6736753}.
Thereby reducing the energy consumption and latency caused by communication (offload the task to MEC server) and computing. 
The advantage is that the task can be cached on the RPN in advance. 
After the task is generated, as long as the UE finds that the task has been cached, it can directly download the corresponding results from the RPN. 
According to the four possible relationships across different task generation process mentioned in Section \ref{Sec02A-1}, it is easy can find that for the same task at different times, if their corresponding data is different, their corresponding result may also be different. There is no point in caching such a task.

In this section, we have examined various research works related to computation offloading. 
A common observation is that the majority of these studies overlook the optimization of task types and often neglect the significance of data in this context. 
While some research on task splitting in computation offloading does consider task type optimization, their primary focus is generally on enhancing the flexibility of resource and task allocation. 
However, these studies tend to overlook the core of tasks, which is data. 
There are only a few studies that recognize the crucial role of data in computation offloading. 
This area of research is still nascent, suggesting that further exploration into data-centric computation offloading is both necessary and promising.

	\section{Related Surveys on Computation Offloading}\label{Sec031}
Many studies related to computation offloading have been published. 
Thus, researchers have surveyed these publications from various perspectives. 
In this paper, we mainly focus on task type and energy consumption to survey the computation offloading related research. 
In addition to our proposed survey perspective, some researchers surveyed computation offloading related work from other viewpoints and perspectives.

Heidari {\em et al}. \cite{heidari2020internet} surveyed computation offloading based on the IoT scenario and proposed a new taxonomy for computation offloading based on offloading decision mechanisms and overall architectures. 
The authors also pointed out the future research direction with the corresponding challenges of computation offloading used in IoT. 
Moreover, De Souza {\em et al}. \cite{9239945} reviewed papers about computation offloading in vehicular environments.

Different types of RPN have their own advantages and disadvantages.
Khan {\em et al}. \cite{6553297} surveyed the work which considered the usage of MCC in computation offloading, and they pointed out the advantage and disadvantages of MCC. 
Mach {\em et al}. \cite{7879258} did the literature survey for  MCC in computation offloading from the perspectives of offloading decision,  resource allocation, and mobility management.
Deshmukh {\em et al}. \cite{7567332}, Chalaemwongwan {\em et al}. \cite{7561437}, and Kumar {\em et al}. \cite{kumar2013survey} focused on the architecture of computation offloading in MCC.
Shi {\em et al}. \cite{7488250} surveyed the work about MEC-based computation offloading. 
Unlike other survey papers, the authors focused on the different use cases, such as video analytics, smart home, smart city. They presented several challenges and opportunities for such use cases. 
Peng {\em et al}. \cite{peng2018survey} surveyed the work about MEC-based computation offloading from the perspective of service adoption and provision. 
Feng {\em et al}. \cite{FENG2022103366} conducted a comprehensive survey on the application, offloading objectives, and offloading approaches of computation offloading in MEC. 
They further analyzed the current challenges and future direction from the perspectives of subtasks dependency and online task requests, server selection, real-time environment perception, and security.
Lin {\em et al}. \cite{8758310} conducted a comprehensive survey for computation offloading with MEC in terms of application partitioning, task allocation, resource management, and distributed execution.
Moreover, some researchers also conducted the survey for fog-based computation offloading \cite{rahimi2020fog,gasmi2021survey}.
Combining different RPNs in computation offloading is also one of the focuses of many researchers, so, Wang {\em et al}. focused on the field of cloud-edge cooperative computation offloading systems, and categorized related papers from the perspective of task type, offloading decision, optimization objective,  mobility, and the type of cooperation \cite{9217600}. 

In addition to the literature review mentioned above, some other literature surveys focus on the optimization method used in computation offloading, and use the technology as the viewpoint. 
For example, Shakarami {\em et al}. \cite{shakarami2020review} surveyed the literature which used the game-theory to optimize the computation offloading process for mobile edge computing.  
Wang {\em et al}. \cite{7883826} did a comprehensive survey for the mobile network architecture and surveyed the mobile edge caching technology used in mobile edge computing. 
Shakarami {\em et al}. \cite{shakarami2020survey} reviewed the papers which focused on machine learning-based computation offloading. 
In their work, the researchers classified machine learning-based computation offloading into three types: supervised learning-based mechanisms, unsupervised learning-based mechanisms, and reinforcement learning-based mechanisms.

Some investigations are carried out based on optimization objectives. 
Wu {\em et al}. \cite{8252700} conducted computation offloading investigations from the perspective of the trade-off between energy consumption and response time. 
Energy saving is also a significant key performance indicator (KPI) of computation offloading, and it is also a key direction that many researchers pay attention to, so Cong {\em et al}. \cite{cong2020survey} surveyed the mobile edge computing from the view of hierarchical energy optimization.

The above survey papers related to computational offloading conducted the survey from various perspectives: some from an architectural standpoint, some based on application scenarios, some through optimization methods, and others focusing on optimization objectives (as shown in Table \ref{tabelI}.). 
However, they all overlook the impact of different types of tasks on computation offloading. This has prompted us to conduct a study based on task types.

\begin{table}[]
	\centering
	\caption{Related survey papers}\label{tabelI}
	\begin{tabular}{|c|l|l|}			
		\hline
		\noalign{\hrule height 0.4pt}
		\multicolumn{2}{|c|}{\textbf{Main Survey Entry Point}} & \multicolumn{1}{c|}{\textbf{Surveys}} \\
		\hline
		\multirow{2}{*}{\textbf{Scenario}} & IoT & [191] \\
		\cline{2-3}
		& Vehicle & [192] \\
		\hline
		\multirow{4}{*}{\textbf{Type of RPN}} & MCC &  [2], [193], [194], [196] \\
		\cline{2-3}
		& MEC &  [4], [126], [197], [198] \\
		\cline{2-3}
		& Fog &  [199], [200] \\
		\cline{2-3}
		& Cloud-edge & [35] \\
		\hline
		\multirow{3}{*}{\makecell{\textbf{Optimization}\\ \textbf{Method}}} & Game-theory & [201] \\
		\cline{2-3}
		& Caching & [3] \\
		\cline{2-3}
		& AI & [202] \\
		\hline
		\multirow{2}{*}{\makecell{\textbf{Optimization}\\ \textbf{Problem}}} & Trade-off  & [203]\ \\
		\cline{2-3}
		& Energy & [133] \\
		\hline
		\noalign{\hrule height 0.4pt}			
	\end{tabular}
\end{table}

\section{Computation Offloading with Static Task}\label{Sec03}
The work related to the computation offloading of static tasks (Section \ref{Sec03}) and the computation offloading of dynamic tasks (Section \ref{Sec04}) are surveyed separately. 
In addition to the task type, it can also be seen from step two in Fig. \ref{Fig03} that the offloading strategy also has a great influence on the algorithm for resource allocation and task allocation. 
Therefore, after classifying the computation offloading according to the task type, it further divides each category based on offloading strategy (i.e., full offloading and partial offloading). 
	
In this section, the research on computation offloading involving static task are divided into two categories, full offloading (\ref{Sec03A}) and partial offloading (\ref{Sec03B}), according to offloading strategy, to conduct surveys separately.
\subsection{Full Offloading}\label{Sec03A}
\subsubsection{Static  Task with Energy Consumption Minimization Problem}
Addressing the problem of minimizing energy consumption while meeting latency requirements in static task based computation offloading,
Zhao {\em et al}. \cite{7938331} studied it based on the static task in multi UE system. 
They jointly considered task allocation, radio resource allocation and computing resource allocation for the UE energy minimization problem.  
In order to solve the problem, the author proposed Reformulation-Linearization-Technique based Branch-and-Bound method.
	
Jointly considering resource allocation and task allocation can improve resource utilization to improve energy saving. 
Still, because the number of tasks for processing has not changed, the performance improvement is limited. 
Liu {\em et al}.  \cite{8588982}  further considered edge caching in computation offloading, and they jointly optimized communication resource, computing resource, and caching contents to minimize the energy consumption of UE while satisfying the UE's latency requirement. 
To solve this optimization problem, which is proven as a mixed-integer non-convex optimization problem, the authors  proposed an iterative algorithm based on the joint application of block coordinate descent and convex optimization.

Expanding on energy consumption considerations, in some studies, the energy consumption of RPN is also necessary to consider.
The entire energy consumption minimization (UE and RPN) problem was considered in \cite{10.1145/2333660.2333724}. 
In their study, authors proposed a game theory-based approach for energy minimization problem. 
According to the rewards and punishments obtained after each iteration, iteratively update the offloading decision until the system reaches the Nash equilibrium.

\subsubsection{Static  Task with Latency Minimization Problem}
The latency minimization problem was studied in the edge-could system by  Wu {\em et al.} \cite{8487354}. 
In their assumption, the cloud server can provide all services, but edge server can only provide part of services. 
They formulated the problem as an integer linear programming model, and they designed a hybrid heuristic method based on genetic algorithms and the simulated annealing selection strategy. Furthermore, regarding to the objective of minimizing latency in cloud-edge architecture, Ren {\em et al}. \cite{8664595} further studied the latency minimization problem under assumptions that each user is associated with an edge server and all tasks have the same type and arrive at the same time. 
They decomposed the weighted sum latency minimization problem into two sub-problem: 
1) minimizing the weighted transmission latency between UE and edge server; 2) minimizing the weighted computing latency of the edge server and the cloud server. 
For the first sub-problem, the Cauchy-Buniakowsky-Schwarz inequality can be used to obtain the optimal solution in closed form. 
For sub-problem two, the authors transformed it into a convex optimization problem. 
After that, Karush–Kuhn–Tucker (KKT) conditions can be used to solve the two sub-problems. 
Energy consumption  was used as a constraint to participate in the optimization of computation offloading in \cite{8487354} and \cite{8664595}. 

\subsubsection{Static  Task with Trade-off Problem}	
Recognizing the significance of balancing energy consumption and latency, the trade-off between energy consumption and latency as a popular optimization objective was considered by Wang {\em et al}. \cite{7870694}.
They proposed a scheme by jointly considering the offloading strategy, interference management, and resource allocation in the computation offloading to minimize the weighted energy consumption and latency.  Unlike some papers, the authors assumed that decision for resource and task allocation is determined by MEC server.
The MEC server used the graph coloring method to allocate the offloading decision and physical resource block. Authors considered the same scheme in \cite{7996857}.

One challenge of computation offloading is communication. 
When the quality of the channel is poor (such as obstacle occlusion), the success rate of offloading will be greatly affected, especially in the case of full offloading. 
Designed for this condition,  as a device deployed at high altitude, UAV can be used as a relay to solve this problem well. 
Moreover, UAV can be seen as a fog node that has a limited computing ability that can help UE to process some tasks \cite{8956055,chen2020multilevel}. 
At the very beginning, UAV was regarded as a relaying node and existed in the computation offloading systems, and the computing ability is ignored by researchers \cite{7572068,8247211}. 
However, as research into computation offloading continues, the computing ability of UAV has also attracted the attention of researchers. 
Yu {\em et al.} \cite{8956055} proposed a UAV-enabled MEC architecture to overcome the problem of the poor channel between  IoT devices and a MEC server. 
They aimed to minimize the weighted latency and energy consumption (UE and UAV). 
In their consideration, the system consisted of a set of UEs (no computing ability), a UAV (low computing ability), and a set of ground edge servers (high computing ability). 
The authors proposed a successive convex approximation algorithm to find a sub-optimal offloading solution for minimizing the weighted sum of the latency and energy consumption of all UEs and UAV by jointly considering the UAV position, communication resource allocation and computing resource allocation. 

Continuing the exploration of trade-off between energy consumption and latency, 
another study about the trade-off between energy consumption and latency was introduced in \cite{8533343}, the computation offloading worked on a multi-cell wireless network, and the authors' research purpose was minimizing the weighted sum of task completion time and energy consumption by jointly optimizing task allocation and resource allocation.  
Since this optimization problem is a mixed-integer nonlinear program, which is difficult to solve, the author decomposed this optimization problem into two sub-problems, including the resource allocation problem when the offloading decision is known, and the task allocation problem when the resource allocation is known. 
Finally, the optimal solution is approached by iteratively solving these two sub-problems and updating the known conditions.

It is also a trade-off between energy and latency issues, but Chen {\em et al}.  \cite{7307234} considered it in mobile edge-cloud computing. 
They studied the multi-UE computation offloading problem with multi-channels interference. 
For solving this trade-off problem, they used game theory (GT) to design a computation offloading model. 
Through continuous iterations, the system can reach the state of Nash equilibrium, and a suitable solution for offloading decision and resource allocation decision can be obtained. 
GT is used to model the interaction between two or more users, and it is a mathematical model \cite{9257921,8270639,8653406}. 
UEs can effectively make decisions based on local observations by using GT in task and resource allocation. 
It has lower complexity, and it is often used in the multi-user computation offloading, but it often yields a suboptimal solution.

\subsection{Partial Offloading}\label{Sec03B}
In terms of partial offloading with static task, most papers on computation offloading are based on multi-task (each UE has multi-tasks).  In partial offloading, it is necessary to consider whether the UE needs to offload tasks and which tasks to offload. 
This is also the main reason why the algorithm of partial offloading is more complicated than the algorithm of full offloading.

\subsubsection{Static Task with Energy Consumption Minimization Problem}
Wu {\em et al}. \cite{8854118} investigated the UE energy minimization problem while guaranteeing the latency requirement for the static task in multi-task scenarios.  
They achieved this by exploring computation offloading through non-orthogonal multiple access (NOMA) and jointly considering task allocation, local computing resource allocation, and NOMA transmission duration. 
The authors proposed an iterative optimization approach that optimizes task allocation (including the offloading decision and offloading order) and other resource allocation to achieve better performance.

\subsubsection{Static Task with Latency Minimization Problem}		
With the development of the data-demanding application, the requirements for the communication of VR are also getting higher and higher. 
To support such applications, Du {\em et al.}\cite{9120235} also investigated the use of THz wireless for MEC computation offloading systems in VR to minimize latency.
Authors jointly took into account the viewport rendering and THz downlink power allocation problem by using the asynchronous advantage actor-critic algorithm, and a method based on deep reinforcement learning was proposed to learn the best viewport rendering position and transmission power control, and to adapt the time-varying characteristics of the wireless channel.

\subsubsection{Static Task with Trade-off Problem}
In \cite{7225153}, Wang {\em et al}. considered the case of multi-UE and multi-cloud computation offloading. 
Each UE has multiple tasks, and the weighted energy consumption and time consumption minimization problem can be formulated as an integer linear programming  problem. The authors also optimized this problem separately for two different cases in their paper. 
For the special case, which is UE has unlimited energy and each task has the same resource requirements, the authors designed a polynomial-time optimal solution based on a weighted bipartite matching problem. 
The authors also proposed a novel heuristic-based algorithm to obtain the binary offloading decisions and the communication resource allocation method for the general case. 
	
Similar to \cite{7225153}, Chen {\em et al}. \cite{7510999} also considered the problem of multi-task in a multi-user system for the energy and latency trade-off problem. 
The optimization problem was modeled as an NP-hard, non-convex, quadratic constrained quadratic programming problem, and the authors proposed a separable semidefinite relaxation with the heuristic algorithm. 
Moreover, since the author considered the problem of multi-task, this will result in an overlap of task processing times and transmission times. 
Still, the authors did not analyze this overlap, they only considered the time lower bound (the degree of overlap in processing time reaches the maximum, only the largest one need to be taken into account) and time upper bound (no overlap in processing time), and use their corresponding performance (obtained by different method) as the lower and upper limits of system performance respectively for the performance benchmarking. 
Building upon \cite{7510999}, as a further extension, Chen {\em et al}. \cite{8438562} further considered the multi-level collaborative computation offloading architecture in the multi-task multi-UE system for energy and latency trade-off problem. 
Based on the modeling in \cite{7510999}, the computing access point (a limited computing node located between UE and cloud) was taken into account. 
For solving this problem, the authors proposed a method called `MUMTO-c', which has a similar principle with the method proposed in \cite{7510999}.

\section{Computation Offloading with Dynamic Task}\label{Sec04}
In this section, we will divide the computation offloading research involving dynamic task into two categories, full offloading and partial offloading, according to offloading strategy, to conduct surveys separately. 
The two types of dynamic task are surveyed in this section, i.e., naturally generated dynamic tasks (type one) and dynamic tasks generated by intervening with static tasks (type two).

\subsection{Full Offloading}\label{Sec04A}
\subsubsection{Type One of Dynamic Task}
The energy consumption minimization while meeting the latency requirement problem was investigated in \cite{6574874}.
Zhang {\em et al}. modeled the MCC-based computation offloading with optimal-energy under the random wireless channel. 
When the task is determined to be offloaded to the MCC server for processing, the data transmission rate is dynamically adjusted according to the current channel state information. 
When the task is determined for processing locally, the working frequency of the CPU is dynamically adjusted according to the immediate processing situation. 
In the study, the authors assumed that the input data size is known, but the CPU cycles requirement to process the corresponding task can be shown as a random number with an empirical distribution. 
Therefore, the processing frequency can be dynamically adjusted according to the distribution of random numbers and the actual processing conditions.

As a further extension  of  work at \cite{6574874}, You {\em et al}. \cite{7442079} further considered the EH problem used in that case. 
In their study, they made the same assumptions for the task in this study as the task in \cite{6574874}. 
The proposed microwave power transfer (MPT) and MCC combined based computation offloading for the passive low-complexity devices, and all the energy for mobile devices came from energy-harvesting. When the task was determined to offload to cloud for processing, it need divide the entire time into two time slots. 
The first time slot need to do MPT for energy collection, and did the computation offloading in time slot two by using the collected energy. 
When the task was determined to be processed locally, it will optimize the CPU frequency, the same idea as mentioned in \cite{6574874}. 
Different from the considerations of many papers, in these two papers, they assumed that they do not know the detailed information of processing tasks such as the CPU cycles requirement of each task, and they assumed that the CPU cycles requirement is a random variable with an empirical distribution. 
The situation they considered is a computation offloading based on task unawareness, which can be defined as the dynamic task based computation offloading.

The studies conducted by \cite{6574874} and \cite{7442079} are categorized under a scenario where clock frequency is optimized continuously during the progression of computing.
This is a type of optimization for computation offloading with dynamic task in the time domain. 
In addition, some researchers consider the impact of queue length in the buffer in previous time slot.
Labidi {\em et al}. \cite{7124703} thought that the number of data packets (task) arrival can be described by Poisson distribution. 
For the energy consumption minimization problem while meeting the latency requirement,  Labidi proposed a deterministic and random offline strategy for a single UE system. 
The dynamic environment of the time-varying channel, wireless resource scheduling and computation load were jointly considered during the offloading process. 
As a further extension, \cite{7348043} extended the single UE system of \cite{7124703} to multi-UE system.

In the direction of energy saving for UE, in addition to latency can be used as a constraint, reliability can also be used as a constraint. 
Liu {\em et al}. \cite{8269175} studied the UE energy consumption minimization problem of computation offloading in a multi-UE, multi-MEC, ultra-reliable low-latency edge computing scenario with the constrain of reliability and latency. 
By comparing with the current system designs which are relying on the average queue length, latency, the authors proposed an approach by imposing a probabilistic constraint on the queue length (probability of exceeding length) and using extreme value theory to deal with extreme events.
In  that paper, task arrival will change over time which is the same as \cite{7442079}, but the authors considered modeling the task arrival by Poisson distribution. 
Considering that this way of task arrival will also cause queue congestion, this also motivated the author to consider the queue problem on the other hand. 
The queuing problem is also a common problem that often needs to be taken into account for dynamic tasks.

In \cite{7541539}, Liu {\em et al}. proposed a scheme to minimize latency. 
This was accomplished by finding the optimal offloading decision strategy based on the application buffer queue status, the available processing capacity at the UE and the MEC service, and the channel characteristics between the UE and the MEC server. 
The authors used  Markov decision process to model this problem, and then proposed an efficient one-dimensional search algorithm to find the optimal task scheduling strategy.

\subsubsection{Type Two of Dynamic Task}
By far, the most considered computation offloading related research for type two of dynamic task is task splitting. 
In \cite{8279411}, Liu {\em et al}. studied the trade-off between the latency and communication reliability in MEC computation offloading for ultra-reliable low latency communication. The authors gave an assumption that the task can be split into multiple subtasks, the granularity of task splitting has arbitrary precision, and sequentially offload each subtask with the entire given channel bandwidth. 
The author designed three algorithms based on heuristic search, reformulation linearization technique, and semi-definite relaxation, respectively, and solved the problem by optimizing edge node candidate selection, offloading ranking, and task allocation.

Taking into account the division of tasks that cannot be arbitrarily split, it is crucial to consider task topology, which pertains to the relationships and dependencies among tasks within a system, when splitting tasks. 
This avoids random splits that could cause confusion or errors and ensures the resulting subdivided tasks maintain coherence and relevance within the broader context of the system.
Zhao {\em et al}. \cite{9419755} addressed the challenge of efficient task offloading in MEC, where tasks have specific service requirements and a dependent order of execution. 
The researchers emphasized the implications of constrained service caching at edge nodes on task offloading decisions, which can result in infeasible decisions or longer completion times. 
To address this issue, the authors defined the problem of offloading dependent tasks with service caching and proves that no constant approximation algorithm exists for this problem. 
The authors then proposed an efficient convex programming based algorithm to solve this problem and a favorite successor based algorithm to solve the special case with a homogeneous MEC.

Similarly, regarding the issue of task topology in computation offloading,  Chen {\em et al}. \cite{9257019} proposed a dependency-aware offloading scheme in MEC that utilizes both edge and remote cloud servers for latency minimization problem. 
The offloading problem is divided into two sub-problems (proved as NP-hard problems), each aiming to minimize the application finishing time under different cooperation modes and task dependency constraints. 
Then, two greedy-based algorithms were designed to solve the two sub-problems that were proven to be NP-hard.
Simulation results demonstrate that the proposed algorithms achieve near-optimal performance and outperform existing benchmark algorithms.

\subsection{Partial Offloading}\label{Sec04B}
\subsubsection{Type One of Dynamic Task}
Liu {\em et al}. extended the full offloading method proposed in \cite{8269175} to study partial offloading in their subsequent work \cite{8638800}.
The authors considered the same one-to-two task splitting as mentioned in \cite{8638800}, and they considered a resource allocation problem based on the ultra-reliable low-latency edge computing system for energy efficiency.  
In convention computation offloading, most of the computation offloading systems were designed based on average-based metrics, which is not suitable to be used in the ultra-reliable low-latency edge computing system (reliability requirement). 
Thus, Liu {\em et al}. proposed a new constraint design approach that is suitable for the ultra-reliable low-latency edge computing system by using the extreme value theory to offset the shortcomings of the design which is based on average-based metrics. 
In addition, the authors used the mobility characteristics of UE , and proposed a dual time scale UE-server association and task computation framework. 
In this regard, taking into account the task queue, the computing power, workload of the server, co-channel interference, and ultra-reliable and low-latency constraints, the authors used the matching theory to associate the UE with the MEC server for a long period of time. Then, given the associated MEC server, perform computation offloading and resource allocation in a short time.

Aside from task splitting, energy harvesting for computational offloading is also a widely researched area for type one dynamic tasks.
The scheme proposed in \cite{7442079} that all the energy of UE comes from the energy-harvesting, and it has great limitations. 
For example, the efficiency of EH is very low, so the energy it provided can not support high-power computing and transmission, which is extremely unfriendly for some computation-intensive tasks. 
Therefore, in \cite{7973020}, the authors considered both the green power collected by EH and the energy of the battery of the mobile device itself. 
They also considered the situation where workload arrival will change over time, and then proposed an efficient resource management algorithm based on reinforcement learning.  
The algorithm instantly learns the best strategy for dynamic workload offloading and edge server configuration to reduce long-term system costs to the lowest.  Foresight and adaptability are supposed as two key points for the designed system.

\subsubsection{Type Two of Dynamic Task}
The problem of saving UE energy using EH and task splitting technology for type two dynamic tasks in computation offloading is investigated in \cite{8515736}.
Zhang {\em et al}. considered a combination of EH and task splitting.  In their assumption, energy-constrained mobile devices harvest energy from ambient radio frequency signals, and the task can be split into two parts at any ratio (local processing part and MCC computing part). Then, jointly considering the  clock frequency, transmission power and offload rate of UE to minimize the energy cost of UE by using the alternative optimization based on the difference between convex function programming and linear programming

Still in an arbitrary task splitting method for type two dynamic task, weighted UE energy consumption minimization problem in multi-UEs is considered in \cite{7842016}.
You {\em et al}. studied the multi-user mobile edge computing offloading system based on time-division multiple access (TDMA), and the task splitting was considered. 
Input data can be split into two part, one for local processing, one for remote processing. 
By using TDMA, it divide time into two time slots, one for transmission or local computing, the other time slot for cloud computing and downloading the task result. 
Moreover, assigning the offloading priority to the UE according to the status of the UE, if the UE has a lower priority, it offloads only a minimum amount of computation tasks to meet the latency requirement. Otherwise, it will offload all computation tasks to MEC. 
Resource utilization has been effectively improved by using the task splitting which changes the static task into dynamic task.  
Orthogonal frequency-division multiple access is also considered as extension work in \cite{7762913}, which has a get a better performance in energy saving, and the energy consumption is reduced by 90\%.

Another code base task splitting for type two dynamic task is investigated in \cite{8603743}.
As a further extension of \cite{8279411}, Liu proposed a computation offloading based on code-partitioning in \cite{8603743}. 
The authors still considered the trade-off problem between latency and reliability, but the difference is that,  the reliability considered by the author was service reliability which was combined with communication reliability, computation-reliability, and the probability that the latency does not exceed the requirements. 
In order to solve this trade-off problem between reliability and latency, the author proposed an algorithm based on integer particle swarm optimization (IPSO). 
Although its result is close to the optimal solution of the problem, the complexity of the algorithm is too high. 
Therefore, the author proposed a heuristic algorithm with lower algorithm complexity but performance similar to IPSO.

Task splitting indeed provides a valuable way to transform static tasks into dynamic ones (type two), resulting in performance improvements. 
However, as mentioned, this approach does not optimize computation offloading from the perspective of input data, and a large amount of redundant data may still be used for data transmission and processing.
Zhang {\em et al}. \cite{9448864} recognized this issue, and they addressed this issue by jointly considering task allocation, resource allocation, and data removal in computation offloading systems to minimize UE energy consumption. 
They observed that in the time and task domains, input data corresponding to most tasks is redundant and detrimental to task processing.
To tackle this problem, the authors proposed a similarity check method in the time domain and the task domain, aiming to remove highly correlated data, reduce redundancy, and improve the UE's energy efficiency.
This approach marks a shift in the focus of computation offloading research, with Zhang {\em et al.} paying more attention to the correlation between source/data in the time domain. 
Similar views can be found in \cite{2103.15924, 2104.03818}, further emphasizing the importance of addressing data redundancy and correlation in computation offloading systems.

\section{Conclusion and Future Computation Offloading}\label{Sec05}
As shown in the previous sections, computation offloading has attracted significant attention in recent years. 
Researchers have studied computation offloading from various perspectives, and their results reflect the superiority of computation offloading in energy saving, latency reduction, and more.
In this section, we will present our conclusion and summarize our understanding of future computation offloading and the corresponding challenges.

In recent years, there has been a significant increase  in research related to computation offloading (see Fig. \ref{Fig01}), as it has gained considerable attention and interest from the scientific community.
According to our survey, these computational offloading studies predominantly concentrate on allocation steps, including resource allocation and task assignment. 
These aspects have been extensively researched and explored in the field for static task based computation offloading.
In the field of dynamic task based computation offloading, type one dynamic tasks are often investigated in scenarios that involve ultra-reliable and low-latency computation offloading requirements, focusing on task queuing problems with extreme theory.
Type two dynamic task are often investigated from the perspective of task splitting with optimum spitting ratio optimization problem.
The splitting-ratio problem for different scenarios/optimization objectives are also well studied.

In computation offloading systems, energy efficiency can be represented by dividing the total payment by the total data size. 
Total payment may include factors such as time cost, energy cost, monetary cost, or others, depending on the optimization objective. 
In conventional computation offloading, the total data size of tasks is a fixed value that cannot be optimized. 
The only thing that can optimize is the payment by using resource allocation, task allocation. 
This also implies that there is a fixed minimum value for the total payment, which cannot be surpassed, regardless of employing techniques such as task splitting, game theory, machine learning, or heuristic algorithms. 
Consequently, when the total payment reaches its upper bound, conventional approaches become ineffective. This defines the upper bound of performance for conventional computation offloading. 

As a result, the type two dynamic task with redundancy removing deserves to be further investigated for computation offloading. 
By focusing on this type of task, researchers can potentially uncover new ways to optimize the total data size and improve energy efficiency, latency reduction, and overall performance in computation offloading systems. 
Exploring redundancy removal methods, such as the similarity check method proposed by Zhang {\em et al}. \cite{9448864}, could lead to breakthroughs that significantly enhance the capabilities of computation offloading and expand its applications.

The aim of redundancy removing is to remove data that has no effect on task processing,  this also involves the co-design of communication system and control signal (result of task).
So, It is believed that the research on computation offloading will turn to computation offloading for communication and control co-design. 
In this regard, the next two points need to be studied:

\begin{itemize}
	\item The modeling for redundancy removing based computation offloading under the stationary data.
	\item The modeling for redundancy removing based computation offloading under the non-stationary data.
\end{itemize}
The data can be divided into two different types, stationary data (the frequency of data change is predictable) and non-stationary data (the frequency of source change is unpredictable). 
For different types of data, their most suitable optimization methods are different. However, the current redundancy removing based work for computation offloading does not pay attention to this because they did not consider the complexity of the redundancy removing algorithm and its corresponding time and energy consumption. 
They only care about energy consumption and latency caused by communication and computing as well as they focus on the computation offloading from the task level. 
They do not make the most of the impact of the data in their work

Partial offloading will significantly increase the complexity of task allocation algorithms, so the corresponding algorithms (resource allocation and task allocation) will consume a lot of energy and cause high latency (although the time and energy cost from this part are ignored by most of the researchers now). 
The complexity problem is often overlooked in computation offloading, and this problem is challenging to solve. 
On the contrary, although full offloading has a lower flexibility perspective, its algorithmic complexity is also lower and it is easier to implement in industry.
Therefore, full offloading will be the most appropriate offloading strategy when complexity issues are not addressed, because of the low algorithmic complexity of full offloading. Furthermore, different data collected by different sensors of the same UE, or data collected by different UEs, may be correlated.
Using full offloading, different UEs can share the data at the RPN  to achieve higher reliability than using only the data collected by the UE itself, or reduce the energy consumption and latency caused repeated computation. 
Moreover, when moving to 6G, THz communication becomes more and more critical. So that how to combine the full offloading with THz communication will be a hot spot research direction.  
In particular, the integrated  and sensing communication \cite{9830717,9851463} under THz communication has the characteristics that sensing and communication can be carried out at the same time, which is very conducive to full offloading and data removing, but the work in this direction has not yet started.

\balance

\bibliographystyle{myIEEEtran}	
\bibliography{ref.bib}

\begin{IEEEbiography}[{\includegraphics[width=1in,height=1.25in,clip]{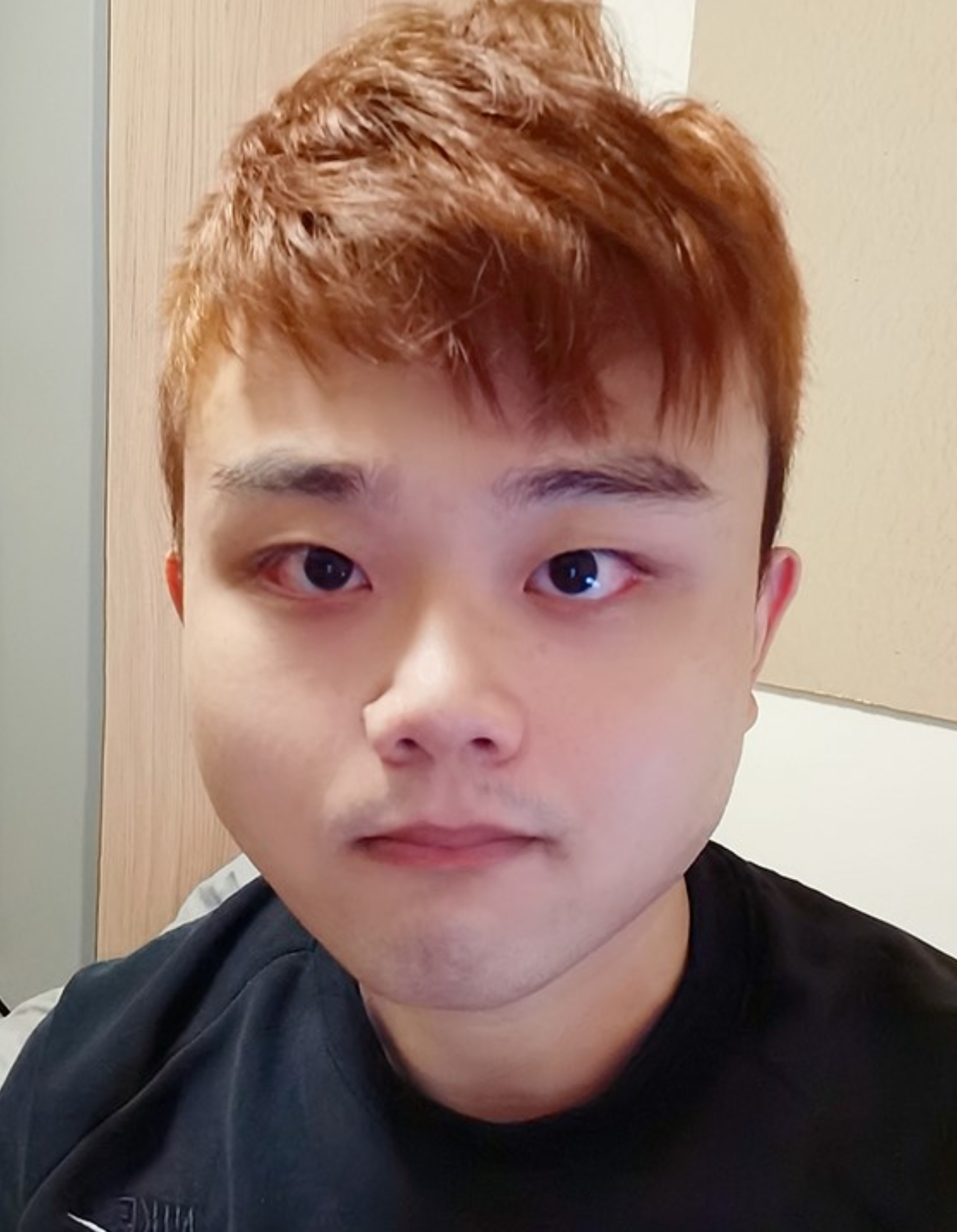}}] {Siqi Zhang} (Graduate Student Member, IEEE) received the B.Sc. degree in electronic information technology from the Macau University of Science and Technology, Macau, China, in 2019, and the M.Sc. degree in wireless communication from the University of Surrey, U.K., in 2020. He is currently pursuing the Ph.D. degree in wireless communication with 5GIC and 6GIC, Institute for Communication Systems (ICS), University of Surrey. His main research interests include edge computing, semantic communication, goal-oriented communication, communication and control co-design.
\end{IEEEbiography}

\begin{IEEEbiography}[{\includegraphics[width=1in,height=1.25in,clip]{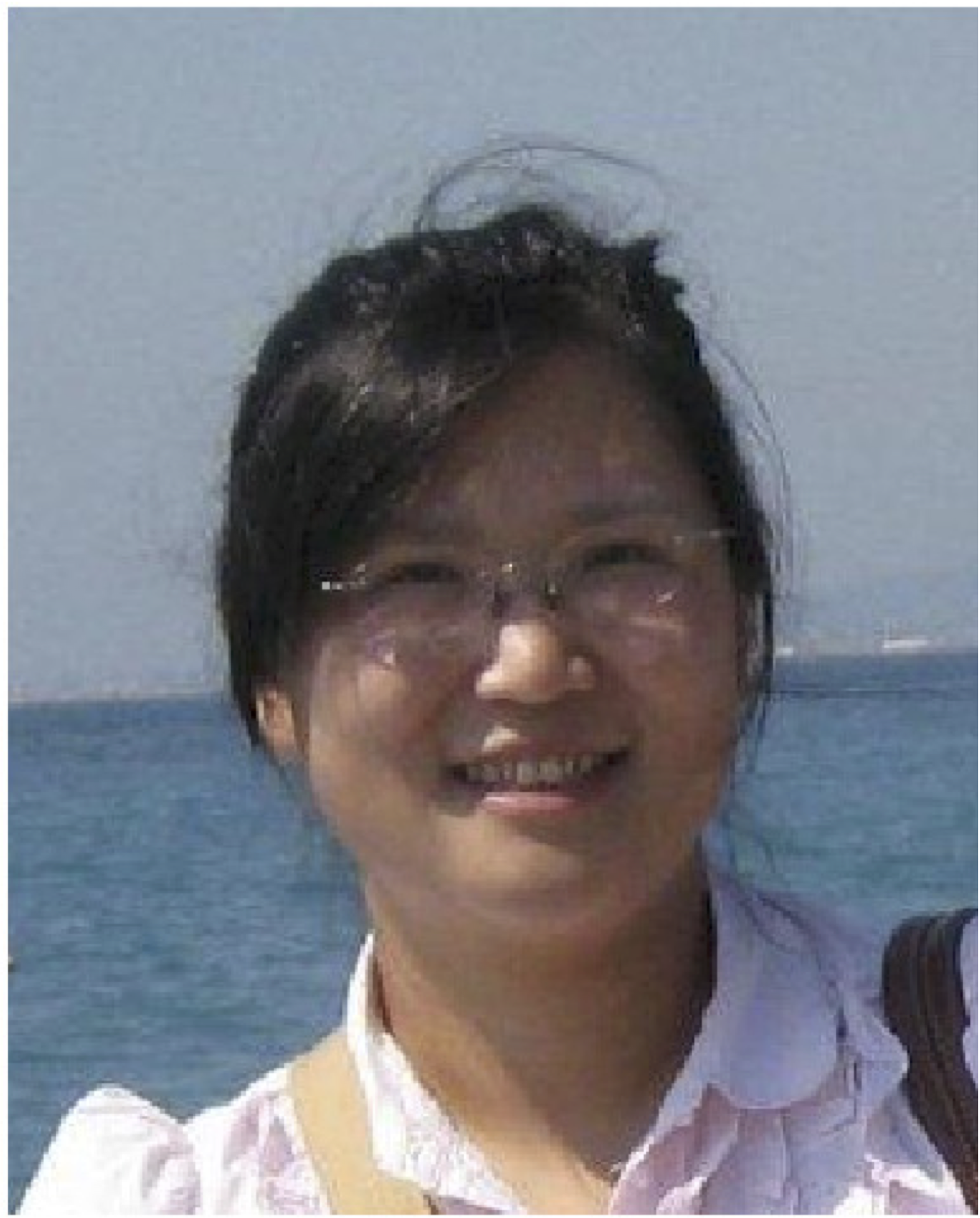}}]{Na Yi} (M'10) received the Ph.D. degree in wireless communications from the University of Surrey, U.K., in 2009. Her research interests and expertise cover cognitive radios/networks, advanced signal processing in wireless communications and sensor networks, cooperation and relaying in wireless networks, and cognitive resource allocation for wireless networks. 
\end{IEEEbiography}

\begin{IEEEbiography}[{\includegraphics[width=1in,height=1.25in,clip]{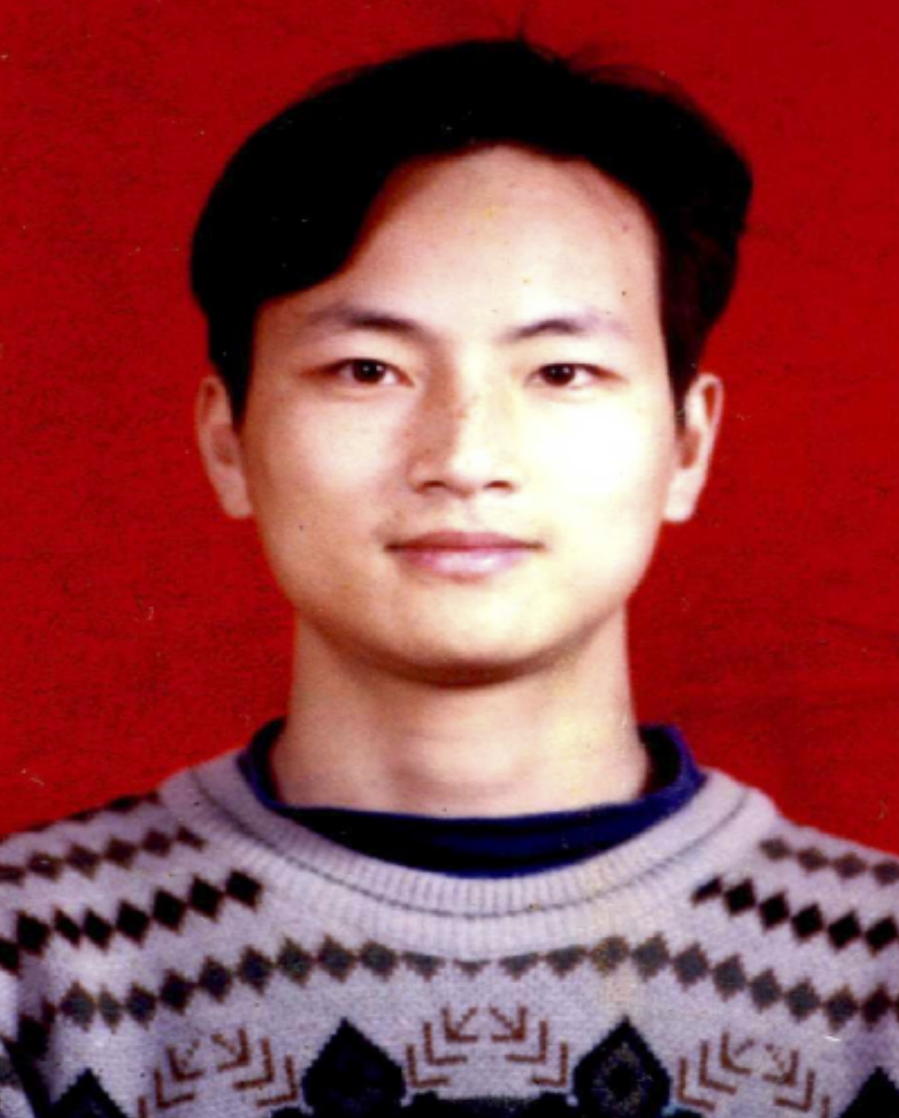}}]{Yi Ma} (Senior Member, IEEE) joined the Institute for Communication Systems (ICS), University of Surrey, Guildford, U.K., in 2004, where he is currently a Chair Professor. He is the Head of the Artificial Intelligence for Wireless Communication Group, ICS, to conduct the fundamental research of wireless communication systems covering signal processing, applied information theory, and artificial intelligence. He is the Work Area Leader of the 5G and 6G Innovation Centre, University of Surrey. He has authored or coauthored more than 150 peer-reviewed IEEE journals and conference papers in the areas of deep learning, cooperative communications, cognitive radios, interference utilization, cooperative localization, radio resource allocation, multiple-input–multiple-output, estimation, synchronization, and modulation and detection techniques. He holds eight international patents in the areas of spectrum sensing and signal modulation and detection. He has served as the Tutorial Chair for EuroWireless2013, PIMRC2014, and CAMAD2015. He is the Chair of the Air-Interface Club, ICS. He is the Co-Chair of the Signal Processing for Communications Symposium in ICC'19. He was the Founder of the Crowd-Net Workshop in conjunction with ICC'15, ICC'16, and ICC'17.
\end{IEEEbiography}

\end{document}